\documentclass[prb,twocolumn,showpacs,superscriptaddress,oneside, floatfix]{revtex4}%
\usepackage{amssymb}
\usepackage{amsfonts}
\usepackage{graphicx}
\usepackage{epsfig}
\usepackage{amsmath}
\begin{document}

\title{High Field Studies of Superconducting Fluctuations in High-$T_{c}$ Cuprates\\ Evidence for
a Small Gap distinct from the Large Pseudogap}
\author{F. Rullier-Albenque}
\email{florence.albenque-rullier@cea.fr}
\affiliation{Service de Physique de l'Etat Condens\'e, Orme des Merisiers, CEA Saclay
(CNRS URA 2464), 91191 Gif sur Yvette cedex, France}
\author{H. Alloul}
\affiliation{Laboratoire de Physique des Solides, UMR CNRS 8502, Universit\'e Paris Sud,
91405 Orsay, France}
\author{G. Rikken}
\affiliation{Laboratoire National des Champs Magn\'{e}tiques Intenses, UPR 3228, CNRS-UJF-UPS-INSA, 31400 Toulouse, France}
\date{\today}

\begin{abstract}
We have used large pulsed magnetic fields up to 60 Tesla to suppress the
contribution of superconducting fluctuations (SCF) to the ab-plane
conductivity above $T_{c}$ in a series of YBa$_{2}$Cu$_{3}$O$_{6+x}$ from the
deep pseudogapped state to slight overdoping. Accurate determinations of
the SCF contribution to the conductivity versus temperature and magnetic field
have been achieved. Their joint quantitative analyses with respect to Nernst data allow us 
to establish that thermal fluctuations following the Ginzburg-Landau
scheme are dominant for nearly optimally doped samples.The deduced coherence 
length $\xi(T)$ is in perfect agreement with a gaussian (Aslamazov-Larkin) contribution 
for $1.01T_{c} \lesssim T \lesssim 1.2T_{c}$. A phase fluctuation contribution
might be invoked for the most underdoped samples in a $T$ range which increases 
when controlled disorder is introduced by electron irradiation. 
For all dopings we evidence that the fluctuations are highly damped when increasing $T$ or $H$.
This behaviour does not follow the Ginzburg-Landau approach which should be
independent of the microscopic specificities of the SC state. The data
permits us to define a field $H_{c}^{\prime}(T)$ and a temperature 
$T_{c}^{\prime}$ above which the SCF are fully suppressed. 
The analysis of the fluctuation magnetoconductance in the Ginzburg-Landau approach 
allows us to determine the critical field $H_{c2}(0)$.
The actual values of $H_{c}^{\prime}(0)$ and $H_{c2}(0)$ are found quite
similar and both increase with hole doping. These depairing fields, which
are directly connected to the magnitude of the SC gap, do
therefore follow the $T_{c}$ variation which is at odds with the sharp
decrease of the pseudogap $T^{\ast}$ with increasing hole doping. This is
on line with our previous evidence that $T^{\ast}$ is not the
onset of pairing. So the large gap seen by
spectroscopic experiments in the underdoped regime has to be associated with the
pseudogap. We finally propose here a three dimensional phase diagram
including a disorder axis, which allows to explain most peculiar
observations done so far on the diverse cuprate families. 
\end{abstract}

\maketitle

\section{Introduction.}

The occurrence of a pseudogap \cite{Alloul89} in the phase diagram of 
high-$T_{c}$ cuprates has raised many questions which are still intensely
debated. Immediately after its discovery, the important issue was to know whether it might be
connected to superconductivity (SC). The fact that the onset temperature $T^{\ast}$ of 
the pseudogap has been found quite robust with disorder contrary to $T_{c}$ 
has been a strong indication that the two phenomena are not directly related \cite{Alloul-PRL91}. 

While this mainly resulted from quasi static NMR measurements in the
1990's, a large amount of new data using energy and/or wave vector resolved
spectroscopies have followed during the last 10 years. 
STM experiments revealed first that  the gap structure 
detected below $T_{c}$ in underdoped samples of Bi$_{2}$Sr$_{2}$CaCu$_{2}$O$_{8}$ (Bi2212) did not totally
disappear above, and transforms into a dip in the density of states \cite{Renner}.
From ARPES experiments, it was found as well that a gap structure is observed in
the normal state of underdoped samples \cite{review-ARPES}. The energy gap detected at low $T$
was found to match the $T_{c}$ variation in overdoped samples, i.e. to
increase continuously with decreasing doping but continues to increase in
the pseudogap state, when $T_{c}$ decreases \cite{Renner}. So putting all these
observations in perspective has justified the idea that the pseudogap could
be a precursor pairing state and not an independent crossover or ordering
occuring in the normal state.

This preformed pair scenario has been strengthened by the observations in
the underdoped cuprates of a large positive Nernst heat transport coefficient
well above $T_{c}$ up to an onset temperature $T_{\nu}$ \cite{Capan,
Wang-PRB2006, FRA-Nernst}. This was  attributed to vortices and/or phase
fluctuations of the superconducting order parameter, along a line
suggested initially from finite-frequency conductivity data \cite{Corson}.
Indeed in these compounds with small superfluid density 
$n_{s}$, it has been proposed that $T_{c}$ is determined by the phase
stiffness of the superconducting order parameter and can be much lower
than the mean field critical temperature $T_{c}^{MF}$ \cite{Emery}. These
experiments, together with the observed diamagnetism above $T_{c}$ 
\cite{Wang-PRL2, Li-PRB2010} have entertained the idea that a precursor pairing
model could be viable to explain the pseudogap. However others suggest
that it could be due to a magnetic order, such as stripes, nematic order or
orbital currents, which could compete or at least interfere with
superconductivity \cite{stripes, STM-nematic, Varma, Bourges}.

In a previous report, based on the experimental approach using high magnetic
fields to suppress the superconducting fluctuatons (SCF), which will be described in
great detail here, we could determine precisely altogether the onset
temperatures $T_{c}^{\prime}$ of the SCF and of the pseudogap $T^{\ast}$
\textit{within the same set of transport experiments} \cite{Alloul-EPL2010}. We have shown there 
that $T^{\ast}$ occurs below
the onset of SCF at optimal doping, demonstrating unambiguously that the
pseudogap cannot be a precursor state for superconducting pairing and has
then to be related to a distinct magnetic order.
  
This pseudogap issue being settled, it remains that these Nernst
experiments evidence that SC pairing extends
above $T_{c}$, which raises important questions about the nature and high $T$
extension of the SCF. Indeed, as for thin superconducting films the SCF
are expected to extend well above $T_{c}$ in 2D systems as compared to the
case of classical 3D BCS superconductors where they disappear in a
vanishingly small $T$ range \cite{Skocpol}. This has initiated a large effort to
study the extension of SCF in thin metallic films. In particular, it has been
demonstrated that a large Nernst signal remains as well above $T_{c}$ \cite{Pourret-NPhys2006}, 
displaying strong similarities on the qualitative aspects with what is observed in the cuprates
\cite{Pourret-NewPhys2009}. Therefore a detailed quantitative study of the SCF in
these systems with  short coherence length $\xi$ is highly
desirable as it might help as well to clear some issues concerning the SC state in high-$T_c$ cuprates.

Since the early days of superconductivity, one of the simplest way to study
SCF has been to determine their effect on the electrical conductivity \cite{Skocpol}. The
fluctuation excess conductivity has been usually well interpreted in the
Ginzburg-Landau (GL) formalism in terms of gaussian amplitude fluctuations of the
order parameter \cite{Larkin-Varlamov}. Among the different contributions
which can be at play, the Aslamazov-Larkin (AL) term either in 2D or 3D
appear the most relevant in high-$T_{c}$ cuprates \cite{SCF-YBCO,
SCF-Bi2212, SCF-Tl, Cimberle}. However in the majority of experiments
reported to date, analyses of the excess conductivity - denoted as
paraconductivity in the AL framework- have been limited to optimally doped
compounds. Indeed in this case, it has been postulated
that the linear $T$ dependence of the resistivity observed in the normal
state can be extrapolated down to low $T$. The SCF contribution to
the conductivity has been then estimated by the deviation from this linear
behaviour. As we shall demonstrate in this work such an assumption
unavoidably introduces large errors if the normal state resistivity deviates
from $T$ linear. Also it is unable to give a reliable estimate of the
highest temperature at which SCF can be detected as this temperature is a
priori imposed by the analysis. Such a criticism is also valid for the magnetoconductance studies in which
the normal state contribution is either totally neglected \cite{Lang, Semba, Holm, Bouquet} or accounted for by an approximative extrapolation from the high-$T$ normal state behavior \cite{Sekirnjak, Ando-PRL2002}.

Recently we have proposed an original method based on the
behavior of the magnetoresistance in high magnetic fields to determine the
field $H_{c}^{\prime}$ and the temperature $T_{c}^{\prime}$ above which
the normal state is completely restored \cite{RA-HF}. 
We have insisted on the fact that $T_{c}^{\prime}$ was indeed a reliable
determination of the onset of SCF. In the present paper we have been able to improve the data
accuracy and to extend the measurements for different hole dopings. This allowed us to 
perform a quantitative analysis of the SCF contribution to the conductivity, and of its $T$ and
$H$ dependence.

After describing the experimental details in section II, we
completely determine the normal state variations of the transport
properties in section III. We obtain then accurate
determinations of the SCF contribution to the conductivity versus $T$
and $H$ (section IV) both for slightly overdoped and underdoped
compounds. The incidence on the SCF of extrinsic controlled disorder
introduced by low $T$ electron irradiation is studied as well.

In section V we give evidence that $T_{c}^{\prime}$ is slightly larger than
the onset $T_{\nu}$ of Nernst effect we have taken before on the same
samples \cite{FRA-Nernst} and that $H_{c}^{\prime}$ is comparable to the
onset field of SCF deduced from Nernst signal or diamagnetic contributions to the magnetization 
\cite{Li-PRB2010}. 

We then take advantage of this unique set of accurate data to
perform a quantitative analysis of the SCF conductivity (section VI). By confronting these
results to Nernst measurements, we do
establish then (section VI.B) that, up to $1.1T_{c}$,
the gaussian AL contribution which decreases as 
$\epsilon =(\ln(T/T_{c})^{-1}$ explains quantitatively both data around optimal doping.
This approach fails in the case of the most underdoped sample, so that contributions of phase 
fluctuations might be invoked there in a small range of temperatures above $T_{c}$ (section VI.C).
Above this range, which increases markedly in presence of disorder, gaussian
amplitude fluctuations of the order parameter again dominate. In any case,
for all the samples studied, we obtain an accurate determination of the $T$
dependence of the coherence length $\xi(T)$ and of its $T=0$ limit.

In Section VII, we show that the analysis of the excess magnetoconductivity
in the Ginzburg-Landau regime allows us to estimate the upper-critical fields
$H_{c2}(0)$ which are found to increase with doping, similarly to the $H_{c}^{\prime}(0)$ 
values. From this observation, we can conclude that the superconducting gap increases with
doping contrary to the pseudogap which decreases.

In section VIII, we study  how the SCF vanish with increasing temperature and magnetic field.
We find for all our samples that the SCF magnitude
drops sharply at high $T$ to vanish near $T_{c}^{\prime}$. We point out that
the cut-off which must be invoked to explain that behaviour implies that the
density of fluctuating pairs vanishes at $T_{c}^{\prime}$. Moreover (section
VIII.B) the field dependence of the SCF conductivity displays a
similar and quite robust exponential dependence in $H^{2}$,
whatever the hole doping or the quantity of disorder. This behavior again
suggests that $T_{c}^{\prime}$ and $H_{c}^{\prime}(T)$ are upper limits
fixed by the vanishing of the pair formation energy.

We then discuss in section IX the results obtained in the present paper in the context of the large 
set of data accumulated on the cuprates in the last decade. We draw there conclusions on various 
important aspects of the normal state and SC properties, and on the incidence of disorder. 
We shall confirm the independence of the pseudogap from the pair formation and give some clues which 
might help to clarify the one gap-two gaps dichotomy in these materials. 

\section{Samples and measurements}

\begin{figure}
\centering
\includegraphics[width=8cm]{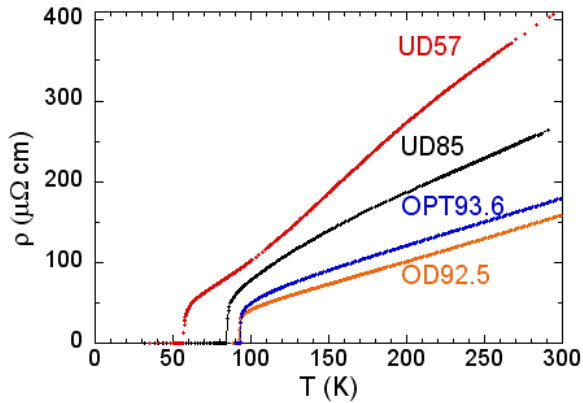}
\caption{(color on line) Temperature dependences of the resistivites of the
four different pure YBCO$_{6+x}$ samples studied.}
\label{Fig.rho curves}
\end{figure}

YBa$_{2}$Cu$_{3}$O$_{6+x}$ (YBCO) single crystals were grown using the flux method. 
Low resistance contacts were achieved by evaporating gold pads in a 
standard four probe geometry. Subsequent annealings in different atmospheres were 
performed in order to get samples with various oxygen contents. We have studied 
four different samples labelled following the values of their critical 
temperatures $T_{c}$ taken at the mid-point of the resistive transition: 
two underdoped samples  UD57 and UD85, an optimally doped sample 
OPT93.6 and a slightly overdoped one OD92.5. The estimate of the hole doping $p$ has been 
done using the parabolic relationship between $T_{c}$ and $p$ \cite{Tallon}. This yields oxygen contents 
of 6.54, 6.8, 6.91 and 6.95 respectively. Although this is not a totally 
secure method \cite{RMP}, it helps at least to proceed comparisons between data on similar samples.
The resistivity curves of the four different samples are displayed in Fig.\ref{Fig.rho curves}

The magnetoresistance (MR) measurements were performed at the LNCMI-Toulouse 
in a pulsed field magnet up to 55-60T. The magnetic field was applied along 
the $c$ axis in order to better suppress SC and its polarity was reversed 
to eliminate by summation any Hall effect contribution to the MR 
determination. 

Controlled disorder was introduced by electron irradiation at low $T$ in optimally doped or underdoped 
samples with $T_{c}\sim57$K. This type of irradiation provides an efficient
way to create point defects, copper and oxygen vacancies, in the CuO$_{2}$ planes, uniformly 
distributed throughout the samples \cite{Legris}. Their effect on the transport and superconducting
properties have been extensively studied previously \cite{FRA-EPL, FRA-PRL2001, FRA-PRL2003}. 
Whatever the hole doping, we have shown that Matthiessen's rule is
well verified at high temperature - as the high $T$ parts of the $\rho(T)$ curves shift parallel 
to each other. This confirms that the hole doping is not significantly modified. This type of 
irradiation results in modifications of the superconducting properties very similar to those obtained 
with Zn substitution \cite{RMP}. In particular, the rate of $T_{c}$ decrease which is around 
- 10K per defect \%-
in the CuO$_{2}$ plane (Zn impurities or Cu vacancies) in optimally doped YBCO$_7$ becomes twice 
larger in underdoped YBCO$_{6.6}$ \cite{Alloul-PRL91, FRA-EPL}

\section{High Field magnetoresistance: Normal state and SC contributions}

\begin{figure}
\centering
\includegraphics[width=8cm]{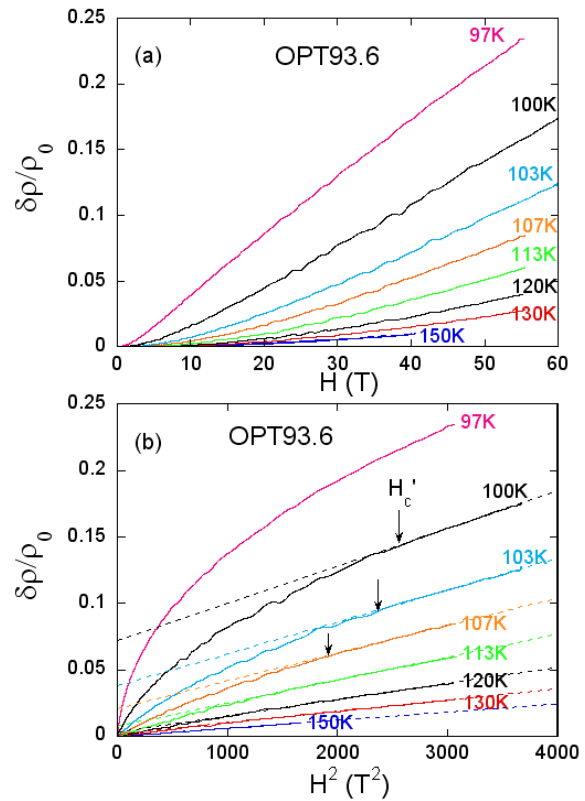}
\caption{(color on line) Resistivity increase normalized to its zero field value plotted versus 
(a) $H$ and (b) $H^{2}$ for temperatures above $T_c$ in the optimally
doped sample OPT93.6. For $T$ $\geqslant140$K, a $H^{2}$ dependence of the magnetoresistance,
represented as dashed lines in (b), is observed for all values of field. For lower $T$, it is 
only seen above a given magnetic field $H^{\prime}_{c}$ (arrows), which is taken as the threshold 
field necessary to completely restore the normal state.}
\label{Fig.magneto-OPT}
\end{figure}

\begin{figure}
\centering
\includegraphics[width=8cm]{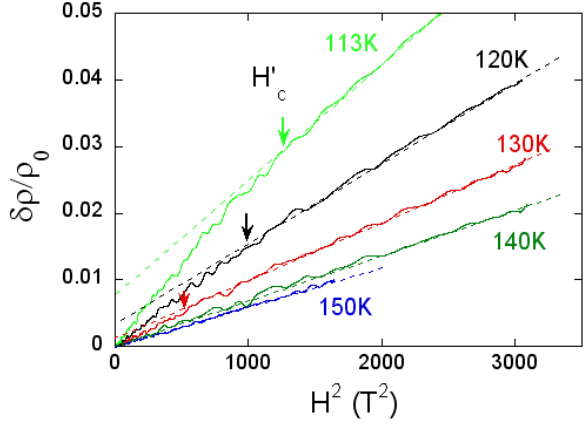}
\caption{(color on line) Enlarged view of the variation of the magnetoresistance in optimally
doped sample OPT93.6 which allows to better visualize the deviations from the $H^2$
normal state dependence for $T \leqslant 130$K.}
\label{Fig.magneto-enlarged}
\end{figure}

\begin{figure}
\centering
\includegraphics[width=8cm]{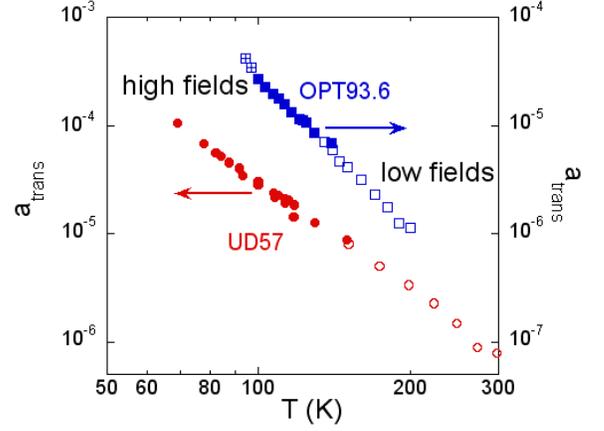}
\caption{(color on line) The magnetoresistance coefficient $a_{trans}$
measured at 55 Tesla (closed symbols) is plotted versus $T$ in logarithmic scales and
compared to that obtained at low field and higher $T$ (empty symbols),
for both optimally doped OPT93.6 and underdoped UD57. Low field data in this latter case are taken from ref.\cite{Harris}. In both cases the
continuity of the data underlines that the magnitude of $H$ has been sufficient to restore 
the normal state.}
\label{Fig.magneto-coeff}
\end{figure}

Fig.\ref{Fig.magneto-OPT} shows the tranverse magnetoresistance (MR) curves measured on the OPT93.6
sample for $T$ ranging from above $T_{c}$ to 150K. 
Similar curves are obtained for all the samples studied. At high $T$, the transverse MR increases 
as $H^{2}$ as better seen in Fig.\ref{Fig.magneto-enlarged}. Such a magnetic field
dependence has been previously observed in different cuprates for $H\leqslant14$T 
\cite{Lacerda,Harris, Konstantinovic}. More precisely in YBCO, Harris et al. \cite{Harris} 
have shown that the weak field magnetoresistance 
$\delta\rho_{n} /\rho_{n0}=(\rho_{n}(H)-\rho_{n}(0))/\rho_{n}(0)$ can be expressed as:
\begin{equation}
\delta \rho_{n}(H)/\rho_{n}(0) = a_{trans}H^{2} \simeq (\omega_{c}\tau_{H})^{2}
\label{eq.magneto}
\end{equation}
where $\omega _{c}=eH/m^{\ast}$ is the cyclotron frequency and $\tau_{H}$ is a transverse 
relaxation time inferred from the Hall angle as $\tan(\Theta_{H})=\omega_{c}\tau_{H}$. 
Let us notice here that Eq.\ref{eq.magneto} refers to the orbital MR coefficient $a_{orb}=a_{trans}-a_{long}$
which would require the knowledge of $a_{long}$, the longitudinal MR . As this latter has been shown to be 
negligible by Harris et al. \cite{Harris}, we have assumed here that $a_{orb}\simeq a_{trans}$.
As Hall constant measurements show that $\cot(\Theta_{H})$ has a quadratic temperature dependence,
this explains the $T^{-4}$ behaviour of $a_{trans}$ observed in ref.\cite{Harris}. The data obtained there 
in weak magnetic fields are displayed as open symbols in Fig.\ref{Fig.magneto-coeff}, for optimally doped 
and underdoped YBCO. At sufficiently high temperature, we also observe a $H^{2}$ variation 
under high magnetic field in our samples. 
This is illustrated by the $H^{2}$ fitting curves in Fig.\ref{Fig.magneto-OPT}(b) 
or in Fig.\ref{Fig.magneto-enlarged} for the OPT93.6 sample. This indicates that the weak field 
limit still applies in OPT93.6 up to 55T. 

However large departures with respect to this 
quadratic behaviour appear when $T$ is lowered towards $T_{c}$. 
The MR steadily evolves from a quadratic to 
a nearly linear field dependence. As stated in ref.\cite{RA-HF}, this evolution can 
be better viewed in the plots versus $H^2$ of fig.\ref{Fig.magneto-OPT}(b)
or Fig.\ref{Fig.magneto-enlarged}. There it can be seen that the $H^{2}$ variation is still visible for 
fields exceeding a $T$ dependent threshold field $H_{c}^{\prime}(T)$, which progressively increases with decreasing $T$.
We attribute the initial faster increase of $\delta\rho/\rho$ 
with $H$ to the destruction of the fluctuating contribution to the conductivity
by the applied magnetic field. In such a case the normal state MR coefficient $a_{trans}$
can then be estimated from the slope of $\delta \rho /\rho$ versus $H^{2}$
at our highest available field (55Tesla). These values of $a_{trans}$ are reported in 
Fig.\ref{Fig.magneto-coeff} 
together with the values determined at low field ($<8$ Tesla) on the same sample for 
$T\geqslant 140K$. We can see there that the data obtained in high field at
low $T$ are in continuity with those obtained at higher $T$, which
underlines the validity of our analysis, and ensures us that we have
effectively completely restored the normal state in high fields for 
$100$K$\leqslant T\leqslant 140$K for the optimally doped sample. However one can notice in 
Fig.\ref{Fig.magneto-coeff} a small upturn of $a_{trans}(T)$ for $T<100$K (crossed squares) 
which signals that it is no longer possible to totally suppress
the superconducting fluctuations even with 55T at 97K, that is 4K above $T_{c}$.

\begin{figure}
\centering
\includegraphics[width=8cm]{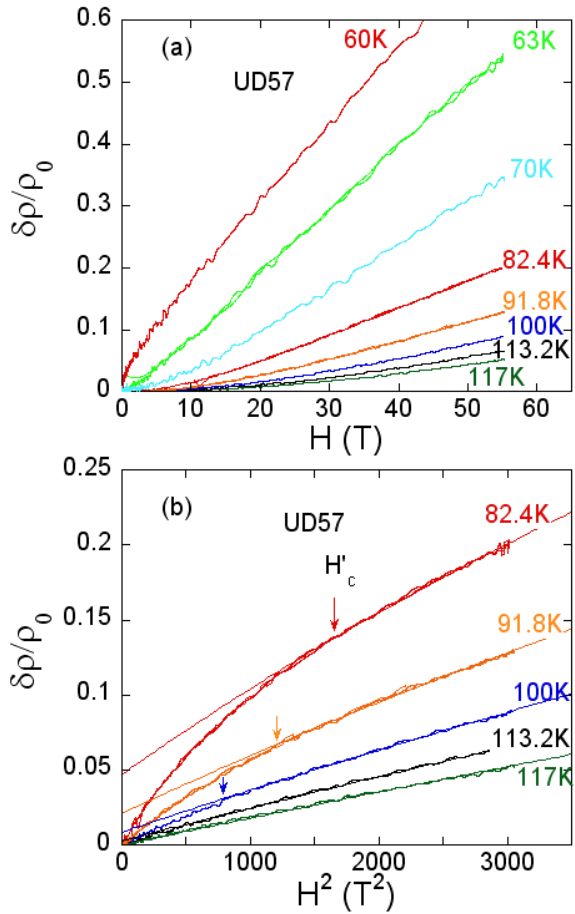}
\caption{(color on line)(a) Field variation of the resistivity 
for decreasing temperatures down to 60K in UD57. (b): The magnetoresistance is plotted
versus $H^{2}$ for $80\leqslant T \leqslant117$K. The full lines are fitting curves using
Eq.\ref{Eq.magneto-sat} in the high field range.}
\label{Fig.magneto-UD}
\end{figure}

Similar analyses have been done for the OD92.5 and UD85 samples.
The $\delta\rho /\rho$ data obtained for the UD57 sample are plotted in Fig.\ref{Fig.magneto-UD} 
versus $H$ or $H^{2}$ in a more limited $T$ range. One can see in Fig.\ref{Fig.magneto-UD}(a) the 
same evolution of the MR as observed for the OPT92.5 sample, from a quadratic to a nearly linear 
field dependence. However, at $\sim 3$K above $T_{c}$, the magnitude of the MR 
is about a factor three larger in the UD57 sample than in the OPT93.6 one. This comes not only
from the larger transverse MR in the normal state but also from an enhanced
contribution of the SCF as we shall see below. 

In such a case one might expect a small saturation of the normal state MR at large field which can 
be expressed as \cite{Tyler-thesis}:
\begin{equation}
\delta\rho_{n} /\rho_{n}=\frac{(\omega_c\tau_H)^2}{1+(\omega_c\tau_H)^2}
\label{Eq.magneto-sat}
\end{equation} 
Eq.\ref{Eq.magneto-sat} gives better fits to the normal state data displayed as full lines 
in Fig\ref{Fig.magneto-UD}(b) for magnetic fields larger than $H_{c}^{\prime}$.
From these fits one can deduce the values for $a_{trans}(T)=(\omega_c\tau_H/H)^2$, which are reported in 
Fig.\ref{Fig.magneto-coeff} \cite{footnoteMR}. They are found here again to be in very good agreement 
with those obtained in low fields and at higher $T$ by Harris \textit{et al} \cite{Harris}.
The low $T$ values of $a_{trans}(T)$, slightly larger than in our previous report \cite{RA-HF} are indeed
consistent with this small saturation of the MR in the UD57 sample. This analysis improves the accuracy 
of the determination of $H_{c}^{\prime}(T)$ without modifying its magnitude compared to our former 
data \cite{RA-HF}. 

Let us finally point out that the behaviour of the normal state magnetoresistance shown in 
Fig.\ref{Fig.magneto-coeff} is quite similar in the optimal and underdoped case nearly down to $T_c$
and exhibits no sign of the reconstruction of Fermi surface 
which is observed at lower $T$ for the YBCO$_{6.6}$ \cite{Doiron-Leyraud, Leboeuf}. 
This point will be discussed in more detail in section IX. 

\section{SCF Contribution to the conductivity}

We have shown here above that it is possible to fully recover the normal state 
conductivity at temperatures slightly above $T_{c}$ when the pulsed field
exceeds $H_{c}^{\prime}(T)$. By extrapolating the normal
state variations of the resistivity $\rho_{n}(T,H)$ down to zero field, one can thus determine the 
normal state value of the resistivity $\rho_{n}(T,H)$ at each temperature and magnetic field. Consequently, assuming only 
that a two fluid model applies, it is straightforward to extract the zero field excess conductivity due to
SCF from
\begin{eqnarray}
\Delta\sigma_{SF}(T,0) & = & \sigma(T,0)-\sigma_{n}(T,0)\, \nonumber \\
& = & \rho^{-1}(T,0)-\rho_{n}^{-1}(T,0)\
\label{para_0}
\end{eqnarray}

In the main studies of superconducting fluctuations in high-$T_c$
cuprates performed up to date, the determination of the fluctuation excess conductivity
has been done in optimally doped samples by assuming that the linear $T$ dependence 
of the normal state resistivity observed at high $T$ can be extrapolated down to low temperature. 
As this assumption can introduce some controversies in the analysis of SCF - and we will show 
below that it is effectively not correct - the study of the fluctuation magnetoconductivity defined as
\begin{equation}
\Delta\sigma(T,H)=\rho^{-1}(T,H)-\rho^{-1}(T,0)
\label{magnetoconductivity}
\end{equation}
has been often preferred since no assumption on the $T$ dependence of the normal state transport 
properties is required in this case \cite{Lang}. Nevertheless, as the corresponding studies have been performed 
in rather weak magnetic fields, it has been always admitted that
the normal state magnetoresistance can be neglected \cite{Lang, Semba, Holm, Bouquet}, i.e. $\sigma_{n}(T,0)\simeq 
\sigma_{n}(T,H)$. 

\begin{figure}
\centering
\includegraphics[width=8cm]{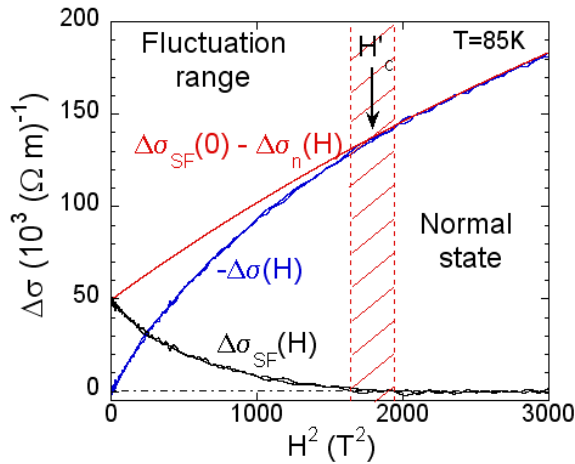}
\caption{(color on line) Decomposition of the magnetoconductivity $-\Delta\sigma(H)$ measured at 85K for the UD57 sample in a normal state contribution and a superconducting contribution. The zero field value of $\Delta\sigma_{SF}(H)$ gives the value of the paraconductivity at 85K.}
\label{Fig.decomposition}
\end{figure}

However for the high magnetic fields used in this study, this assumption is 
not valid and the normal state magnetoconductivity has to be taken into account to deduce 
the field variation of the SCF contribution \cite{Sekirnjak, Ando-PRL2002}. Within a two fluid model, we simply write
\begin{eqnarray}
\Delta\sigma_{SF}(T,H) & = & \sigma (T,H)-\sigma _{n}(T,H)\, \nonumber \\
& = & \rho^{-1}(T,H)-\rho_{n}^{-1}(T,H)\
\label{Def-SFconductivity(H)}
\end{eqnarray}
From the relations above the measured total variation of the conductivity is therefore
\begin{equation}
\Delta\sigma_{SF}(T,H)= \Delta\sigma(T,H) + \Delta\sigma_{SF}(T,0) - \Delta\sigma_{n}(T,H)
\label{def-magnetoconductivity}
\end{equation}
where the normal state conductivity is
\begin{equation}
\Delta\sigma_{n}(T,H)= \rho_{n}^{-1}(T,H)-\rho_{n}^{-1}(T,0).
\label{def-normal_magneto}
\end{equation}
This decomposition, which allows us to obtain $\Delta\sigma_{SF}(T,H)$ is illustrated in 
Fig.\ref{Fig.decomposition} for magnetoresistance data taken at a fixed temperature
$T=85$K in UD57. 

It is worth to emphasize here that the method developed 
in the present work allows us to determine unambiguously the normal-state contribution 
in the presence or absence of magnetic field. We have thus been able to analyse separately 
the variation of $\Delta\sigma_{SF}(T,0)$ with $T$ and that of $\Delta\sigma_{SF}(T,H)$ 
with $H$ at each $T$ for different hole dopings.

\subsection{Zero field excess conductivity versus T: onset temperature $T_c^{\prime}$}

\begin{figure}
\centering
\includegraphics[width=8cm]{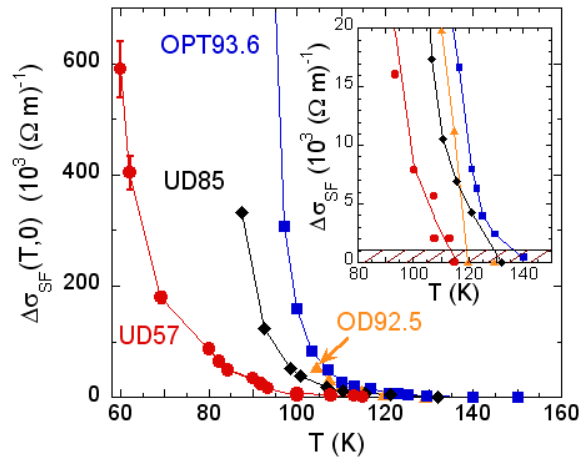}
\caption{(color on line) SC fluctuation contribution to the zero field
conductivity $\Delta\sigma_{SF}(T,0)$ for the four YBCO samples studied here \cite{footnotedatalowT}. 
The enlargement of the high $T$ range shown in the inset
gives an estimate of the accuracy on the determination of $T_{c}^{\prime}$,
the onset of SC fluctuations. Lines are guides for the eyes.}
\label{Fig.paraconductivity-pure}
\end{figure}

Let us first consider the $T$ dependences of the zero field
excess conductivities $\Delta_{SF}(T,0)$ which are reported in Fig.\ref{Fig.paraconductivity-pure} 
for the four pure samples considered here. One can notice that this quantity dies out very
fast with increasing $T$, which allows us to define an onset temperature $T_{c}^{\prime}$. Given the noise
level of the experiments, we have chosen as in ref.\cite{Alloul-EPL2010} to define $T_{c}^{\prime}$ 
as the temperature where $\Delta\sigma_{SF}(0)$ is lower than 1 $10^{3}(\Omega.m)^{-1}$, 
as indicated in the inset of fig.\ref{Fig.paraconductivity-pure}.
Let us note that the decrease in $\Delta_{SF}(T,0)$ with temperature is much slower for the most 
underdoped sample than for the other ones, so that the extension in temperature of the SCF is larger 
in this sample than for the other ones. 

\begin{figure}
\centering
\includegraphics[width=8cm]{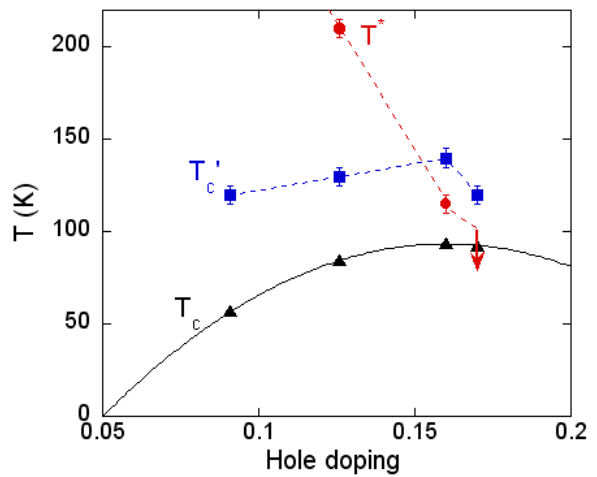}
\caption{(color on line) The values of $T_{c}^{\prime}$ (squares) and $T^{\star}$ (circles) are plotted versus the hole doping for the four samples studied. The full line indicates the superconducting dome. Contrary to $T_{c}^{\prime}$ that is rather insensitive to hole doping, $T^{\star}$ is found to increase with decreasing doping and crosses the $T_{c}^{\prime}$ line near optimal doping \cite{Alloul-EPL2010}.}
\label{Fig.variation-doping}
\end{figure}

The variation of $T_{c}^{\prime}$ with doping is reported in Fig.\ref{Fig.variation-doping}.
The interesting point is that $T_{c}^{\prime}$ is only sightly dependent on hole 
doping and is maximal for optimal doping. We have also reported in this figure the dependence
of the pseudogap temperature $T^{\star}$ whose determination has been done simultaneously using the 
same series of experimental data \cite{Alloul-EPL2010}. The fact that the 
$T_c^{\prime}$ line crosses the pseudogap line near optimal doping shows unambiguously that 
the pseudogap phase cannot be a precursor state for the superconducting phase. 

The quasi-insensitivity of $T_{c}^{\prime}$ to doping observed here
appears very different from what is observed by Nernst or magnetization measurements in single 
layer materials such as La$_{1-x}$Sr$_x$CuO$_4$ (LSCO) or La-doped Bi2201 for which the onset temperature of SCF 
is strongly dependent on hole doping with a sharp maximum in the underdoped region 
\cite{Li-PRB2010}. We will discuss this point in the discussion section IX in conjunction with 
the effect of controlled disorder.

\subsection{Field variation of the SF conductivity: onset field $H_c^{\prime}$ }

\begin{figure}
\centering
\includegraphics[width=8cm]{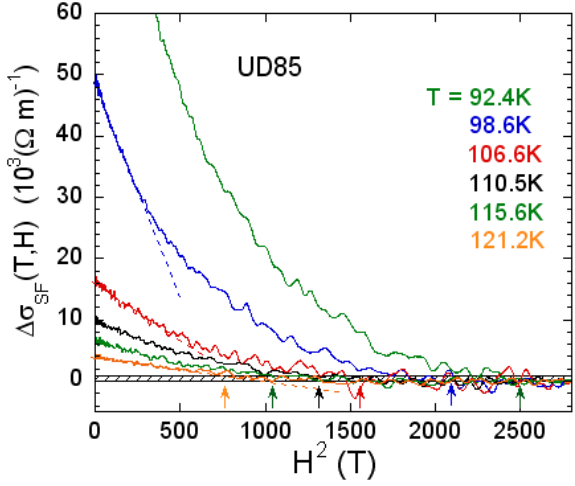}
\caption{(color on line) SC fluctuation contribution to the conductivity $\Delta\sigma_{SF}(T,H)$
in UD85 plotted versus $H^{2}$. The initial linear decays displayed as dashed lines visualize the
quadratic field dependence observed for low magnetic fields. The arrows indicate the
threshold fields $H^{\prime}_{c}(T)$ taken at $\Delta\sigma_{SF}(T,H)=1\,10^3 (\Omega.m)^{-1}$.}
\label{Fig.field-variation}
\end{figure}

In the same way the variation of $\Delta\sigma_{SF}(T,H)$ with magnetic field allows 
us to analyse how the excess conductivity is destroyed by the applied field. This is exemplified in 
Fig.\ref{Fig.field-variation} for UD85 at $90 \leq T \leq 120$K. We can see that $\Delta\sigma_{SF}(T,H)$
starts to decrease quadratically with $H$ whatever $T$. This $H^{2}$ dependence is clearly visible at 
the highest $T$ for fields up to 30T. The same behaviour is observed 
for all the samples studied in the small temperature range which can be explored by our method. One can 
also notice that the accuracy on $\Delta\sigma_{SF}(T,H)$ decreases with increasing field. This is
due both to an increase of the noise induced by the stresses on the magnet
at the highest field values and a much reduced data acquisition time in the
high field range. This fixes our noise level at about $1\,10^3 (\Omega.m)^{-1}$ at high fields.

We can thus determine the fields $H_{c}^{\prime}(T)$ at
which $\Delta\sigma$ becomes smaller than this value. As $T$ decreases, it becomes difficult 
to ascertain that the normal state is fully reached when $H_{c}^{\prime}(T)$ becomes comparable 
to the highest available field. This makes difficult to deduce precise values of 
$H_{c}^{\prime}(T)$ when they become larger than $\sim45$T.

\begin{figure}
\centering
\includegraphics[width=8cm]{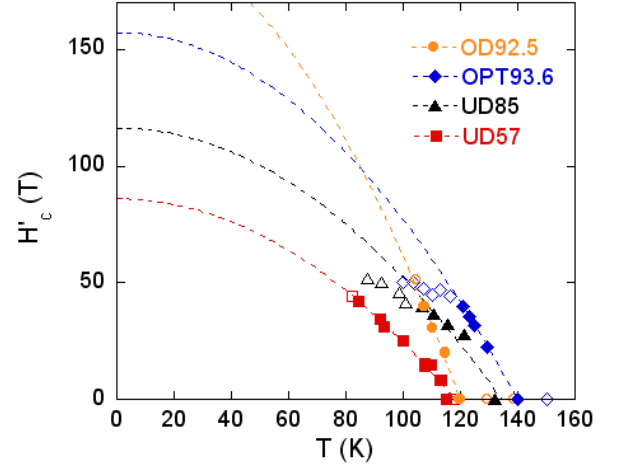}
\caption{(color on line) The field $H_{c}^{\prime}$ at which SC
fluctuations disappear and normal state is fully restored is reported versus 
$T$ for the four pure samples studied. Dashed lines represent the fitting curves 
to Eq.\ref{Eq.H'c-T} using data with closed symbols. When $H_{c}^{\prime}(T)\gtrsim 40T$ (empty symbols), 
the data are somewhat underestimated  as the  maximum applied field is not sufficient.}
\label{Fig.H'c(T)}
\end{figure}

One can see in Fig.\ref{Fig.H'c(T)} that $H_{c}^{\prime}(T)$ drops rapidly with
increasing $T$. For all the samples the variation of $H_{c}^{\prime}(T)$
appears linear near $T_{c}^{\prime}$. It is then tempting to use a parabolic $T$
variation to fit the data as applied for the critical field of classical superconductors. 
\begin{equation}
H_{c}^{\prime }(T)=H_{c}^{\prime }(0)[1-(T/T^{\prime }_{c})^{2}]
\label{Eq.H'c-T}
\end{equation}
The fitting curves displayed as dashed lines in Fig.\ref{Fig.H'c(T)} give 
correspondingly an indication of the field $H_{c}^{\prime}(0)$
required to completely suppress the SC fluctuation contribution down to 0K. It is clear that $H_{c}^{\prime}(0)$ \textit{increases} with hole doping and reaches a value as high as $\sim 150$ Tesla at optimal doping. 

\subsection{Influence of disorder.}
\begin{figure}
\centering
\includegraphics[width=8cm]{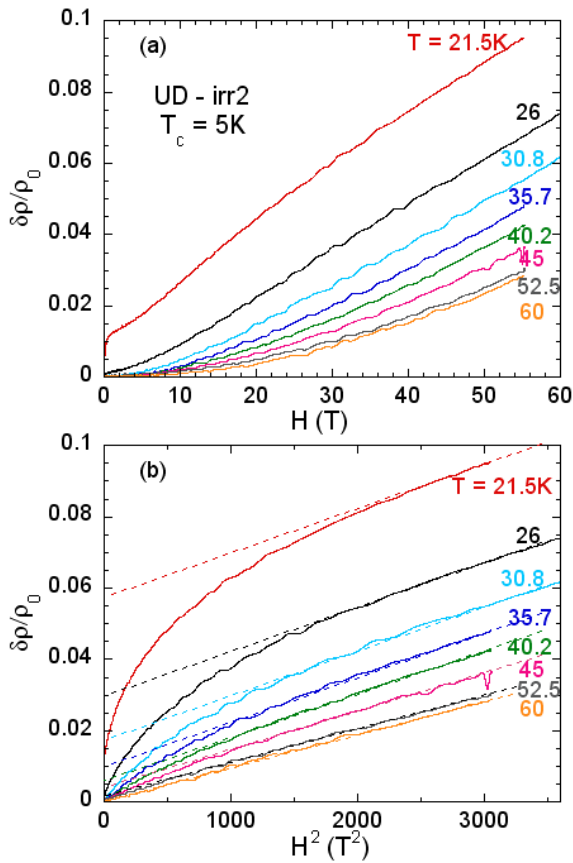}
\caption{(color on line) Magnetoresistivity $\delta\rho/\rho_{0}$ plotted versus $H$ (a) or $H^{2}$ (b) for
a UD57 irradiated sample with $T_{c}=5$K. Dotted lines give the normal state behaviour restored in high fields. They parallel each other, which points out that the magnetoresistance and the scattering time $\tau$ are nearly $T$ independent for strong disorder.}
\label{Fig.magneto-UDirr5}
\end{figure}

In cuprates it is now well established that non magnetic impurity substitutions
or in plane disorder are detrimental to superconductivity and strongly
depress $T_{c}$, the temperature of establishment of 3D superconductivity \cite{RMP}.
However it has been shown as well that the SCF, as seen by Nernst
effect, remain at temperatures much higher than the 3D $T_{c}$ in disordered
samples \cite{FRA-Nernst}. So, we expect to detect paraconductivity contributions well above 
$T_{c}$ in presence of disorder. Let us also notice that increasing disorder decreases markedly
$\omega_{c}\tau_{H}$, which ensures that a $H^{2}$ dependence up to 55T is now perfectly verified 
for the normal state magnetoresistance of irradiated underdoped samples. The magnetoresistivity curves
obtained in an UD57 sample in which $T_{c}$ has been decreased down to 5K are reported in 
Fig.\ref{Fig.magneto-UDirr5} versus $H$ or $H^{2}$.

It is striking to see that even in this low-$T_{c}$ sample, a magnetic field larger than 40T is 
still necessary to totally suppress the superconducting fluctuations at 21K, that is 16K above $T_{c}$.
Moreover we observe that the $H^{2}$ term is nearly $T$ independent, which might be 
quite reasonable in this highly disordered sample for which the relaxation time is expected to become 
rather $T$ independent. This confirms that the normal state behavior is totally restored above the threshold field $H_{c}^{\prime}$.

\begin{figure}
\centering
\includegraphics[width=8cm]{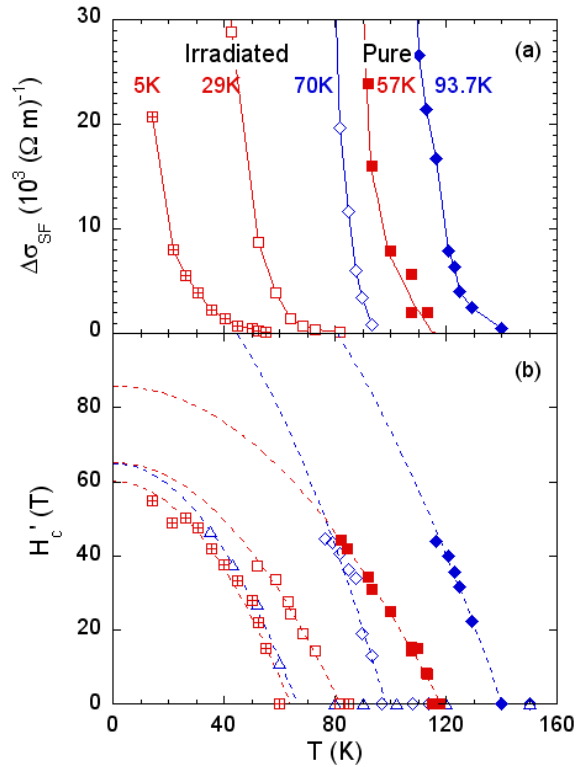}
\caption{(color on line) Comparison of the SCF conductivity $\Delta\sigma_{SF}(T,0)$ in (a) 
and onset field $H_{c}^{\prime}(T)$ in (b) for
pure (full symbols) and disordered (empty symbols) samples of OPT93.6 (diamonds, triangles) 
and UD57 (squares), with the reduced 
$T_{c}$ values as indicated in (a). In (b) dashed lines are
the fitting curves using Eq.\ref{Eq.H'c-T}. Data corresponding to an irradiated OPT sample 
with $T_{c}=30$K (empty triangles) has been added \cite{FRA-EPL-MIT}.}
\label{Fig.results-irr}
\end{figure}

The values of $\Delta\sigma(T,0)$ are reported in Fig.\ref{Fig.results-irr}(a) for OPT and UD57 samples 
either pure or irradiated by electron irradiation at low temperature. We notice that 
the measured SC fluctuation
conductivity $\Delta\sigma$ remains of the same order of magnitude as that
of the pure samples in both cases. Using the same procedure as described above, 
we can also determine $H_{c}^{\prime}(T)$ for the different samples. The corresponding 
values as well as the fitting curves using Eq.\ref{Eq.H'c-T} are displayed in fig.\ref{Fig.results-irr}(b).

\begin{figure}
\centering
\includegraphics[width=8cm]{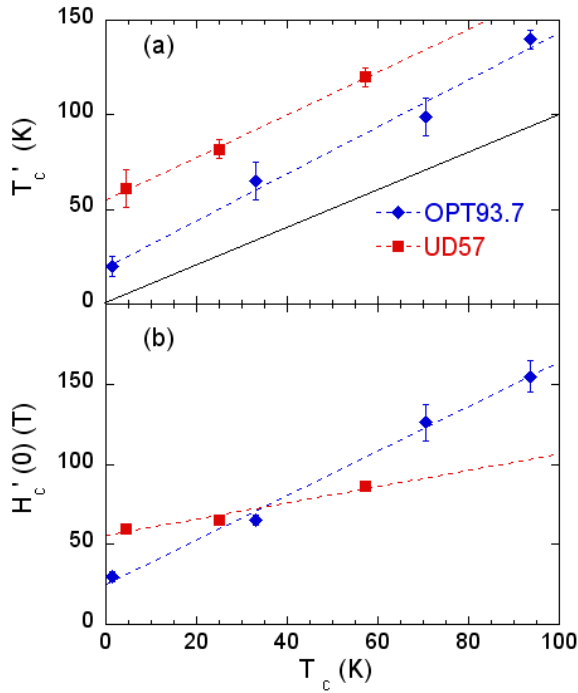}
\caption{(color on line) Variations of (a) $T_{c}^{\prime}$ and (b) $H_{c}^{\prime}(0)$ for OPT (diamonds) 
and UD (squares) pure and disordered samples versus $T_{c}$. Here the $T_{c}$ value is used to monitor the disorder. The full line in (a) corresponds to a slope unity which parallels the  $T_{c}$ variation. Data from ref.\cite{FRA-EPL-MIT} corresponding to irradiated OPT samples with $T_{c}=30$ 
and $1.9$K have been added.}
\label{Fig.T'c-H'c-versusTc-irr}
\end{figure}

In the case of the most irradiated UD sample with $T_{c}=5$K, we have been able to 
nearly completely suppress SC with 55T. The fact that $H_{c}^{\prime}(T)$ can be rather 
well fitted by Eq.\ref{Eq.H'c-T} somehow validates the use of this equation to fit our other data. 
In Fig.\ref{Fig.T'c-H'c-versusTc-irr} where the variations of $T_{c}^{\prime}$ and 
$H_{c}^{\prime}(0)$ are plotted versus $T_{c}$, one can see that both quantities decrease 
with increasing disorder. The reduction in $T_{c}^{\prime}$ nearly follows that in $T_{c}$ for 
the underdoped sample while it is slightly larger for the OPT sample. Consequently, when $T_{c}$ is decreased 
by disorder, \textit{the relative range of SCF with respect to the value of $T_{c}$ expands considerably}.
We also observe that $H_{c}^{\prime}(0)$ decreases linearly with decreasing $T_{c}$, but more
rapidly for the OPT samples than for the UD57 ones. In both cases, even for samples with very low $T_{c}$, magnetic fields 
as large as 30-60T are still necessary to reach $H_{c}^{\prime}(0)$, as can be seen in Fig.\ref{Fig.T'c-H'c-versusTc-irr}-b.

\section{Comparison with results obtained by different experiments}
We have shown that high field resistivity measurements above $T_{c}$ allow us to determine
the temperature range as well as the extension in magnetic field of the fluctuation excess conductivity. 
It is thus very interesting to compare our results with those obtained by Nernst effect or magnetic
torque measurements that have been developed for more than ten years to probe the presence of 
SCF above $T_{c}$ in high-$T_{c}$ cuprates \cite{Li-PRB2010}.
\subsection{Onset temperature for SCF}
The temperature ranges of SCF found here are analogous to those measured in pure high-$T_{c}$ cuprates by 
these other techniques. For optimally doped YBCO, $T_{c}^{\prime}=135(5)$K
found here is in remarkable agreement with the onset obtained from the diamagnetic response in high 
fields \cite{Li-PRB2010}. The variation with doping is also very similar to that observed in Bi2212  
by Nernst effect or diamagnetic measurements \cite{Wang-PRL2}. Even though a very small increase of the 
onset temperature (from 125 to 130K ) is found upon underdoping in this latter case, contrary to our 
results which show a small decrease (from 135 to 120K) in the same doping range, \textit{the important 
point is that $T_{c}^{\prime}$ is not found to vary much with doping, contrary to the pseudogap 
temperature $T^{*}$}.

\begin{figure}
\centering
\includegraphics[width=8cm]{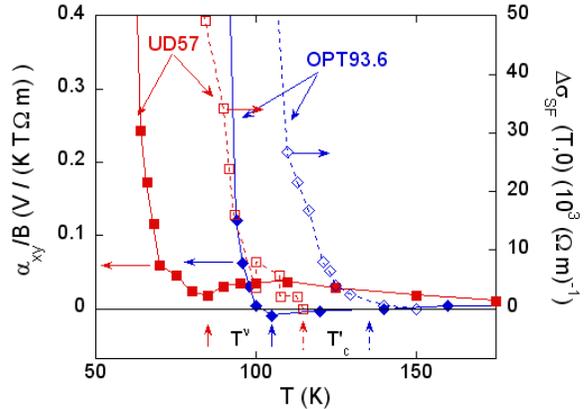}
\caption{(color on line) Comparaison of the onset temperature for SCF extracted from Nernst measurements
($T_{\nu}$) and from this work ($T_{c}^{\prime}$) in the same OPT93.6 and UD57 samples. The closed symbols
are for the off-diagonal Peltier conductivity deduced from the analysis of the Nernst coefficient 
\cite{FRA-Nernst} while the empty ones are the values of $\Delta\sigma_{SF}(T,0)$ obtained in this work. 
The values of $T_{\nu}$ (vertical arrows) had been estimated at temperatures corresponding to 
the minimal values of $\alpha_{xy}/B$.}
\label{Fig.compareNernst}
\end{figure}

It is worth noting that the values of $T_{c}^{\prime}$ found here for the pure compounds are larger than
the onsets of Nernst signal measured previously on the same samples \cite{FRA-Nernst}.
However one can see in Fig.\ref{Fig.compareNernst}, where the determinations of $T_{\nu}$ 
and $T_{c}^{\prime}$ are compared for the OPT93.6 and UD57 samples, that the criterion used to 
deduce $T_{c}^{\prime}$ is much more precise than that for $T_{\nu}$. Indeed a negative $T$ dependent
contribution of the normal state Nernst signal yields a minimum in $\alpha_{xy}/B$, and hides the real onset 
of SCF. It thus appears that the Nernst effect is not the best probe to detect SCF in the case of YBCO.

In a general way, the measured onset marks the point at which instruments lose sensitivity to 
detect superconducting fluctuations. This can explain results of a recent report in which the 
fluctuation excess conductivity measured by Josephson effect between an optimally doped YBCO and 
an underdoped one with $T_{c}=61$K, drops very fast and is found to vanish at $\backsim 15$K above 
$T_{c}$ \cite{Bergeal}. In our case, the excess conductivity of the UD57 sample is still 5\% of its 
normal state contribution at 85K, that is 23K above $T_{c}$. Such an explanation might also 
account for the much smaller fluctuation range determined recently by microwave absorption
measurements in YBCO or mercury compounds \cite{Grbic-YBCO, Grbic-Hg}.

We observe that the gap between $T_{\nu}$ and $T_{c}^{\prime}$ is progressively reduced when disorder 
is introduced in the samples. This is due to the fact that the Nernst signal of normal quasi-particles 
scales inversely with the scattering rate and thus progressively vanishes with disorder, which permits a more
accurate determination of $T_{\nu}$. This results in similar values of $T_{\nu}$ 
and $T_{c}^{\prime}$ in the most irradiated samples. 

\subsection{Magnetic field}
It is also interesting to compare the $H_{c}^{\prime}$ values found here to
the maximum magnetic field $H^{max}$ necessary to completely suppress the Nernst or the diamagnetic 
signals \cite{Wang-PRB2006, Wang-PRL2}. Although early reports have argued that 
$H^{max}$ inferred from Nernst measurements steeply increases with underdoping in Bi2212 \cite{Wang-science},
more recent studies have shown that this field rather increases with doping in agreement with what we 
find here for $H_{c}^{\prime}(0)$. In particular very large values of $H^{max}$ have been estimated from 
torque magnetometry in optimally doped Bi2212 \cite{Wang-PRL2}. At $T_c$, $H^{max}$ is equal to 90T and is thus quite comparable to the value $H_{c}^{\prime}(T_{c})= 87$T deduced here for the OPT93.6 
sample from the fitting curve displayed in Fig.\ref{Fig.H'c(T)}.

One can point out that both values of $H_{c}^{\prime}(0)$ and $H^{max}$ 
are not determined directly by experiments. Here the $H_{c}^{\prime}(T)$ line is 
only accessible above $T_{c}$ and the value of $H_{c}^{\prime}(0)$ is obtained 
using Eq.\ref{Eq.H'c-T}. For the Nernst 
(or magnetization) measurements, the $H^{max}$ values can be only deduced below $T_{c}$ 
by taking the extrapolated field at which the Nernst (or diamagnetic) signal should vanish. 
The fact that similar values of $H^{max}$ and $H_{c}^{\prime}(0)$ are observed in optimally doped Bi2212 
and YBCO gives some support to these two determinations. 

\begin{figure}
\centering
\includegraphics[width=6cm]{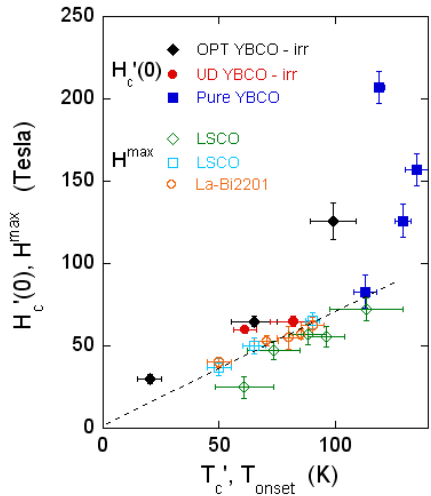}
\caption{(color on line) The maximum fields required to completely suppress
superconductivity at $T=0$, as inferred from Nernst, diamagnetism 
or transport measurements, are plotted versus the onset of superconducting fluctuations
for LSCO samples (empty diamonds \cite{Wang-PRB2006}, empty squares \cite{Li-M2S}), La-Bi2201 
(empty circles \cite{Li-M2S}) and YBCO pure or irradiated crystals (this work). 
While the results for LSCO, La-Bi2201 and irradiated 
YBCO samples follow more or less the same linear dependence (dashed line), the $H_{c}^{\prime}(0)$ values for the pure UD85, OPT93.6 and OD92.7 are much larger with respect to their $T_{c}^{\prime}$ }
\label{Fig.compH'c-Hc2}
\end{figure}

More generally a linear variation of 
these field values versus $T_{c}$  or $T_{c}^{\prime}$ is found when comparing results obtained 
in low $T_{c}$ materials like LSCO or La-Bi2201 and our irradiated optimally doped or underdoped YBCO 
samples as illustrated in Fig.\ref{Fig.compH'c-Hc2}.
As already proposed \cite{FRA-Nernst, RA-HF}, this relationship between $T_{onset}$(or $T_{c}^{\prime}$) and $H^{max}$(or $H_{c}^{\prime}$) leads us to speculate that the presence of defects, either intrinsically present or intentionally introduced by irradiation will play here a significant role. It is worth noting that the parameters found for the pure UD57 sample obey the same quasi-linear relationship. 
However, the results obtained for our other pure samples differ markedly from this behaviour as 
larger values of $H_{c}^{\prime}$ with respect to their $T_{c}^{\prime}$ are found, in agreement 
with reported $H^{max}$ value for OP Bi2212 \cite{Wang-PRL2}. This is of course directly evidenced in 
Fig.\ref{Fig.variation-doping} and Fig.\ref{Fig.H'c(T)} which show that $H_{c}^{\prime}$ increases with hole 
doping while $T_{c}^{\prime}$ remains essentially the same. 

The observation of very large values of $H^{max}$ well above $T_{c}$ has been taken as 
the sign that superconducting fluctuations in all high-$T_{c}$ cuprates originate from vortex-like
excitations in a phase disordered superconductor, rather than fluctuating Cooper pairs \cite{Li-PRB2010}.
This appears today as an overstatement. Indeed recent experiments have evidenced that 
the Nernst signal in NbSi films can be explained solely in terms of gaussian fluctuations even in 
magnetic fields much larger than the orbital upper critical field $H_{c2}$ 
\cite{Pourret-NPhys2006, Pourret-PRB2007}.
Moreover, it has been suggested that the Nernst signal of these films could share some resemblances 
with those seen in cuprates \cite{Pourret-NewPhys2009}. It is thus very interesting to compare 
more quantitatively the evolution of the excess fluctuation conductivity with temperature 
and magnetic field. 

\section{Quantitative analysis of the superconducting fluctuations}
\begin{figure}
\centering
\includegraphics[width=8cm]{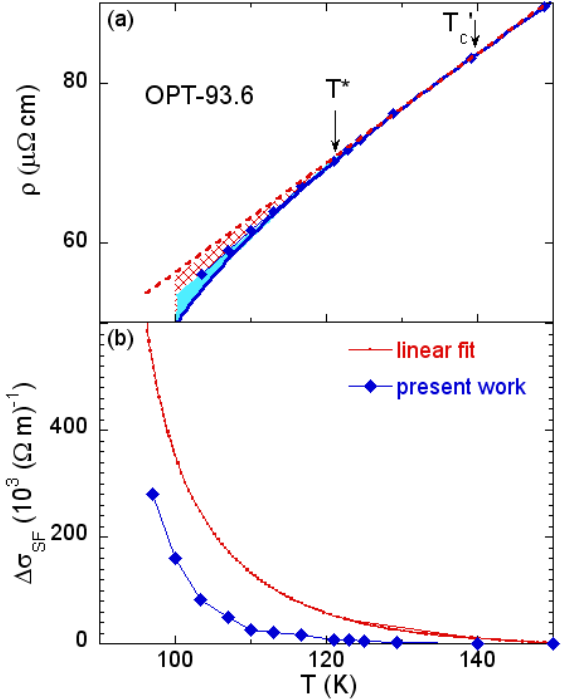}
\caption{(color on line) (a) T variations of the zero-field resistivity (full line)
and the normal state values (symbols) deduced from high-field data for OPT93.6. The coloured 
area corresponds to the range where superconducting fluctuations are effectively present while 
the hatched area is the extra range found by neglecting the decay of normal state resistivity due 
to the pseudogap. (b) The variation $\Delta\sigma_{SF}(T)$ deduced assuming a linear fit would appear 
more accurate though it overestimates the actual value by at least a factor 3.}
\label{Fig.compare-OPT}
\end{figure}

As already pointed out, a lot of studies have been dealing with 
the $T$ dependence of the paraconductivity in optimally doped high $T_{c}$ cuprates \cite{}. But in 
most experiments the magnitude of $\Delta\sigma_{SF}$ deduced from the data are critically dependent on 
the behaviour taken for the normal state resistivity, which has been most often taken as linear in $T$. 
So, we first emphasize here that our method is particularly adapted to perform a precise quantitative analysis of the excess conductivity since our experimental approach allows us to deduce $\sigma_{N}(T)$
reliably. We have for instance shown in our previous work
\cite{Alloul-EPL2010} that the normal state resistivity of the OPT93.6 sample deviates from the 
linear $T$ dependence at $T^{*}\simeq120$K due to the opening of the pseudogap. Consequently, the use of
a linear extrapolation for the normal state resistivity would lead to a large overestimate of 
$\Delta\sigma_{SF}(0)$. To illustrate that, we have therefore mimicked in 
Fig.\ref{Fig.compare-OPT} the difference generated by such an analysis with respect to our 
reliable determination using magnetoresistance data. It is clear that for such a sample the overestimate of $\Delta\sigma_{SF}$ can be quite important. The reliability of the determinations done so far using linear extrapolations of the normal state conductivity can then be put into question in many cases.
 
\subsection{Contribution of gaussian fluctuations to paraconductivity}
The excess fluctuating conductivity is usually analysed in 
the framework of gaussian fluctuations using the Aslamazov-Larkin theory. As for the Maki-Thompson contribution, which is an indirect 
contribution arising from the decay of superconducting pairs into quasi-particles and vice-versa, it 
can be neglected in high-$T_{c}$ cuprates due to strong pair-breaking effects \cite{Yip}. 
In the Ginzburg-Landau theory, the gaussian fluctuations come from the temporal and spatial fluctuations of 
the superconducting order parameter. The corresponding paraconductivity is directly related to the 
temperature dependence of $\xi(T)$, the superconducting correlation length of the short-lived Cooper 
pairs. Upon cooling down to $T_{c}$, $\xi$ is expected to diverge with a power-law dependence given by:
\begin{equation}
\xi(T)=\xi(0)/\sqrt{\epsilon}
\label{def xi(T)}
\end{equation}
where $\xi(0)$ is the zero-temperature coherence length and $\epsilon = \ln (T/T_{c}) \simeq (T-T_{c})/T_{c}$ for $T \gtrsim T_{c}$.
Depending on the relative values of the temperature dependent perpendicular coherence length $\xi_{c}(T)$ 
and of the layer spacing $s$, the paraconductivity can evolve from a 3D behavior in the immediate vicinity 
of $T_{c}$ towards a 2D behavior at larger temperatures \cite{SCF-YBCO}. The paraconductivity can 
be expressed more generally by using the Lawrence-Doniach (LD) theory of layered superconductors as \cite{Lawrence-Doniach}:
\begin{equation}
\Delta\sigma^{LD}(T)=\frac{e^{2}}{16\hslash s} \frac{1}{\epsilon \sqrt{1+2\alpha}}
\label{Eq.LD}
\end{equation}
where the coupling parameter $\alpha = 2 (\xi_{c}(T)/s)^{2}$ with $\xi_{c}(T)=\xi_{c}(0)/ \sqrt{\epsilon}$.
Sufficiently far from $T_{c}$, one expects $\xi_{c}(T) \ll s$ and Eq.\ref{Eq.LD} reduces to the 
well-known 2D Aslamazov-Larkin expression: 
\begin{equation}
\Delta\sigma^{AL}(T)=\frac{e^{2}}{16\hslash s} \epsilon^{-1}=\frac{e^{2}}{16\hslash s}\frac{\xi^{2}(T)}{\xi^{2}(0)}
\label{Eq.AL}
\end{equation}
The only parameters in this expression are the value of the interlayer distance $s$ and the 
value taken for $T_c$ which can have a huge incidence on the shape of the curve especially 
for $(T-T_{c})/T_{c}<0.01$. 

\begin{figure}
\centering
\includegraphics[width=8cm]{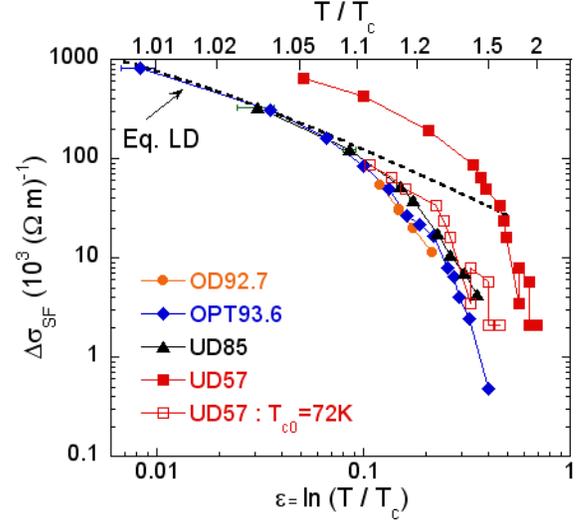}
\caption{(color on line) Superconducting fluctuation conductivity $\Delta 
\sigma_{SF}$ for the four pure samples considered here plotted versus
$\epsilon=ln(T/T_{c})$. Values of $T_{c}$ have been taken here at the midpoint of the resistive transition, and error bars for $\epsilon$ using the onset and offset values of $T_{c}$ are indicated. The dashed line represents the expression of Eq.(\ref{Eq.LD})
with $s=11.7$\AA. and $\xi_{c}(0) \simeq 0.9$\AA . Full lines are guides to the eye.}
\label{Fig.LD-pure}
\end{figure}

We have plotted the variation 
of $\Delta\sigma_{SF}(T)$ for the four different hole contents as a function of $\epsilon$ 
in Fig.\ref{Fig.LD-pure}.
For all the samples except UD57, it is striking to see that our experimental data 
collapse on a single curve. Moreover we can see that the data can be fitted 
reasonably well by the Lawrence-Doniach expression (Eq.\ref{Eq.LD}) in the small temperature range 
$0.03 \leq \epsilon \leq 0.1$ if one takes $\xi_{c}(0) \simeq 0.9$\AA . We have assumed here, as usually
done, that the CuO$_{2}$ bilayer constitutes the basic two-dimensional unit, and $s$ is then taken as 
the unit-cell size in the $c$ direction: $s=11.7$\AA . This is a 
strong indication that \textit{the excess fluctuation conductivity is mainly due to gaussian fluctuations 
in these different compounds}. One can see that all the curves in Fig.\ref{Fig.LD-pure} bend downwards very
rapidly for $\epsilon \gtrsim 0.1$. This behavior has been pointed out 
earlier in different fluctuation studies on YBCO \cite{Freitas, Hopfengartner, Gauzi}. It has 
been proposed that this could be due to the limitations of the Ginzburg-Landau theory in these compounds
with very short coherence lengths. We will discuss this point in more details in paragraph VIII.A. 

It is clear that the situation is completely different for the UD57 sample for 
which $\Delta\sigma_{SF}$ is found to be about a factor four larger than for the other dopings. 
This points to an additional origin of SCF in this underdoped sample. In fact we find that 
the data for this sample can be reconciled with the unique curve found for the other samples 
by using an effective value $T_{c0}$ different from the actual $T_{c}$. This is illustrated in
fig.\ref{Fig.LD-pure} in which the $\Delta\sigma_{SF}$ data of UD57 are also reported versus 
$\epsilon=ln T/T_{c0}$ using $T_{c0}$=72K which is much larger than the actual $T_{c}=57.1$K. 

The same conclusion has been proposed in ref.\cite{Ussishkin} to account for the Nernst signal at 
high $T$ in underdoped LSCO which was found too large to be explained only 
by Gaussian fluctuations.
In the phase fluctuations scenario proposed by Emery and Kivelson \cite{Emery}, 
this would mean that the actual $T_{c}$ is suppressed from the mean-field transition temperature $T_c^{MF}$
by phase fluctuations. However gaussian fluctuations are still expected above this temperature. Thus 
it appears reasonable here to assimilate our effective $T_{c0}$ to $T_c^{MF}$. 
Let us notice that this conclusion is in contrast with that argued from paraconductivity measurements 
in underdoped LSCO samples, in which a description in terms of a 2D AL approach has been proposed 
to completely account for the experimental data \cite{Leridon-PRB2007}

\begin{figure}
\centering
\includegraphics[width=8cm]{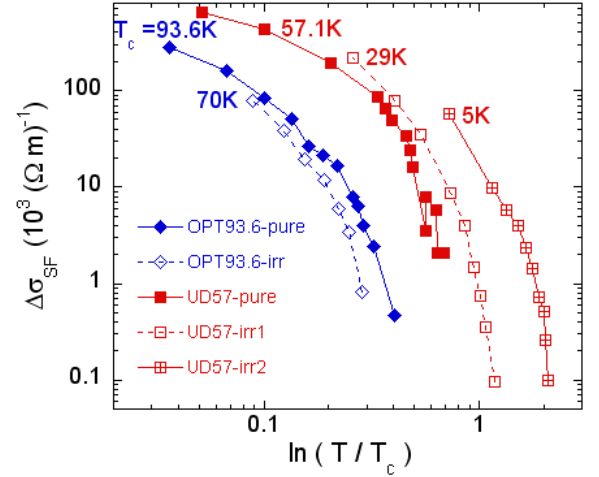}
\caption{(color on line) Same as Fig.\ref{Fig.LD-pure} for the pure and irradiated optimally doped and underdoped YBCO$_{6.6}$ crystals. Full symbols are for the pure samples while empty ones are for the irradiated ones. The shift to the right observed for the UD57 samples with increasing disorder is shown in 
section VI-C to result from the reduction in $T_c$ with respect to the mean field temperature $T_{c0}$.}
\label{Fig.AL-irr}
\end{figure}

It is also interesting to consider the effect of disorder on the paraconductivity. 
As seen in Fig.\ref{Fig.AL-irr}, the curve found for the disordered optimally doped sample with $T_{c}=70$K
nearly falls on that of the pure sample, indicating that here again it is possible to explain the SCF 
in the framework of the AL theory. However, this is not the case for the underdoped samples, 
since their curves are shifted towards larger values of $\epsilon$ with increasing disorder. 
So the introduction of disorder appears to accentuate the difference with the behavior expected from 
a GL approach. This will appear more clearly in the next paragraph. 

\subsection{Nernst effect}
In view of the results found above for the paraconductivity, we have found important to analyse as well our 
Nernst results taken on similar samples \cite{FRA-Nernst} along the same lines. The evolution 
of the off-diagonal Peltier term $\alpha_{xy}$ \cite{footnote-Nernst} has already been recalled in Fig.\ref{Fig.compareNernst} 
for the pure OPT and UD57 samples. 
In the 2D Ginzburg Landau approach, $\alpha_{xy}$ has been found to follow the simple expression \cite{Ussishkin}:
\begin{equation}
\frac{\alpha_{xy}}{B}=\frac{k_{B}e^{2}}{6 \pi \hslash^{2} s} \xi(T)^{2}
\label{Eq.Nernst-gaussianSCF}
\end{equation}
that shows that $\alpha_{xy}$ is related to the spacing between the layers $s$ and the 
Ginzburg-Landau coherence length in the ab-plane $\xi(T)$. Consequently, according to Eq.\ref{Eq.AL}, 
there is a simple linear relationship between $\alpha_{xy}$ and $\Delta\sigma_{SF}(T)$: 
\begin{equation}
\frac{\alpha_{xy}}{B}=\frac{8k_{B}}{3\pi \hslash}\xi(0)^{2} \Delta\sigma_{SF}(T)
\label{Eq.Nernst-sigma}
\end{equation}
whose slope provides a direct determination of the zero temperature coherence length $\xi(0)$. 
Let us point out that, had we taken the LD term in Eq.\ref{Eq.Nernst-gaussianSCF}, it would have been eliminated as the spacing $s$ between layers
in this expression \ref{Eq.Nernst-sigma}.

\begin{figure}
\centering
\includegraphics[width=8cm]{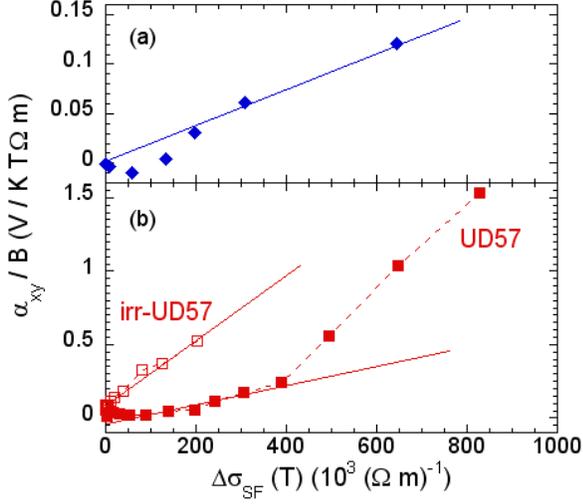}
\caption{(color on line) The values of $\alpha_{xy}/B$ taken from ref.\cite{FRA-Nernst} are plotted 
versus the values of $\Delta\sigma_{SF}(T)$ determined in this work for (a) the pure OPT sample  
and (b) two UD57 samples pure and irradiated. The values of $\Delta\sigma_{SF}(T)$ reported 
here have been obtained by interpolation between the data reported in Fig.\ref{Fig.LD-pure}. Full lines are
linear fits for the high $T$ values while the dashed line is guide to the eye for the data near $T_c$.} 
\label{Fig.Nernst-sigma}
\end{figure}

We have tested this relationship first for the optimally doped case and the
data are plotted in Fig.\ref{Fig.Nernst-sigma}(a). At high $T$ a negative
normal state contribution to the Nernst signal \cite{FRA-Nernst}, apparent
in Fig.\ref{Fig.compareNernst} (similar to that seen by Daou et al. \cite{Daou}) dominates that
due to SCF. Nevertheless near $T_{c}$, this negative counterpart is overcome by the sharp increase
of the positive SCF contribution so that the linear relation of Eq.\ref{Eq.Nernst-sigma} is reliably verified, as can be seen in Fig.\ref{Fig.Nernst-sigma}(a). The linear slope found there near  $T_{c}$  results in a value  $\xi (0)\simeq 1.4$nm after correction for the estimated small 
negative Nernst contribution.

Using the relationship between $H_{c2}(0)$ and $\xi(0)$ 
\begin{equation}
H_{c2}(0)=\Phi _{0}/2\pi \xi (0)^{2},  \label{Eq.H_c2(0)}
\end{equation}
this would lead $H_{c2}(0)\simeq 160$Tesla, a value which resembles that of $H_{c}^{\prime}(0)$ 
determined above. Consequently, we can conclude that, in
optimally doped YBCO, \textit{the paraconductivity and the Nernst signal
above $T_{c}$ are consistent with each other and can be interpreted in terms
of gaussian fluctuations only}, with $\xi (0)\simeq 1.4$nm. 

For the UD57 samples, pure and irradiated with $T_{c}\simeq30$K, the
corresponding data are plotted in Fig.\ref{Fig.Nernst-sigma}(b). The same
analysis can be done for the pure UD57 sample in the high $T$ range (above $T_{c0}$, that is low
values of $\Delta \sigma _{SF}$) where a gaussian regime is expected to be
restored. As can be seen in Fig.\ref{Fig.compareNernst}, the normal state contribution of the
off-Peltier term is positive in this case, so the linear fit of the raw data will give a 
slightly overestimated value $\xi(0)=2.0\pm 0.2$nm. This would 
correspond to $H_{c2}(0)\simeq 80\pm 20$T,
similar here again to $H_{c}^{\prime }\simeq 80$T estimated for
this compound. 

This conclusion is clearly no longer valid at lower
temperatures, near  $T_{c}$, as the SCF contribution of the off-diagonal Peltier term $\alpha_{xy}/B$ 
markedly increases with respect to this linear relation. This confirms our
previous conclusion that the Nernst data in the pure UD57 sample are not
consistent with gaussian fluctuations for $T_{c}<T<T_{c}+15$K. We
furthermore evidence that, in this range, the positive contributions to the
Nernst signal are greatly enhanced with respect to the SCF paraconductivity.

For the irradiated UD57 sample, the quite good linear relationship between
$\alpha_{xy}/B$ and $\Delta\sigma_{SF}$ in fig.\ref{Fig.Nernst-sigma}-b would correspond to 
$\xi(0) \simeq 4.3$nm, hence to a value of $H_{c2}(0)$ as low as 17T while $H_{c}^{\prime}(0) \simeq 60$T 
for this compound. This unrealistic value of $H_{c2}(0)$ implies that Eq.\ref{Eq.Nernst-sigma} 
does not apply in this case.
 
This demonstrates that  the enhancement of the positive Nernst signal by disorder we evidenced in 
ref.\cite{FRA-Nernst} is much larger than that of the paraconductivity. This result contradicts the 
analysis done in ref.\cite{Kokanovic} of the Nernst signals in Zn substituted YBCO thin films.

\subsection{Contribution of phase fluctuations}
As an explanation in terms of gaussian fluctuations is not sufficient to account for the excess conductivity 
in the UD57 samples, either pure or disordered, it is natural to address the possible role of phase
fluctuations. Indeed in these systems with low carrier density and/or high level of disorder \cite{Emery},
the low superfluid density is expected to lead to phase fluctuations below the mean-field 
transition temperature $T_{c}^{MF}$. It has been proposed that the superconducting transition is caused by 
the proliferation of vortices which destroy long-range phase coherence similarly to 
that predicted for the Kosterlitz-Thouless transition (KT) \cite{Lee-RMP}. This would result 
in a phase-incoherent state with a finite pairing amplitude between $T_{c}$ and $T_{c}^{MF}$.
In the framework of the 2D KT transition, the excess conductivity is expressed as:
\begin{equation}
\frac{\Delta\sigma}{\sigma_{n}} \equiv \left(\frac{\xi(T)}{\xi(0)}\right)^{2}
\end{equation}
where the coherence length $\xi(T)$ is now related to the vortex density $n_{v}$ through 
$2\pi n_{v}\equiv 1/\xi^{2}$. It turns out that a similar relation holds also in the AL regime as shown above 
(see Eq.\ref{Eq.AL}), but here $\xi(T)$ is expected to diverge exponentially at $T_{KT}$. 
An interpolation formula between these two regimes has been proposed initially by Halperin and 
Nelson \cite{Halperin-Nelson}. More recently, Benfatto et al. \cite{Benfatto} have revisited this 
problem by means of a renormalization group (RG) approach and have established a direct 
correspondence between the parameter values used to describe the BKT fluctuation regime and 
the reduced temperature $\tau_{c}$ between $T_{KT}$ and $T_{c}^{MF}$ defined by:
\begin{equation}
\tau  \equiv \frac{T-T_{KT}}{T_{KT}} \;\quad \tau_{c} \equiv \frac{T_{c}^{MF}-T_{KT}}{T_{KT}}
\label{Eq.def}
\end{equation} 
They propose an interpolation formula for $T \gtrsim T_{KT}$ which is formally similar to that of 
Halperin and Nelson:
\begin{equation}
\Delta\sigma_{SF}/\sigma_{n} = \left(\frac{2}{A}\right)^{2} \sinh ^{2}\left(\frac{b}{\sqrt{\tau}}\right),
\label{Eq.formulaKT}
\end{equation}
but where the parameters $A$ and $b$ are now obtained from the numerical RG calculations of the correlation length near the transition, so that $A$ is close to unity and $b$ given by 
$b \sim 2\alpha^{\prime}\sqrt{\tau_{c}}$ where $\alpha^{\prime}$ measures the deviation of the vortex 
core energy with respect to the conventional value in the XY model.

\begin{figure}
\centering
\includegraphics[width=8cm]{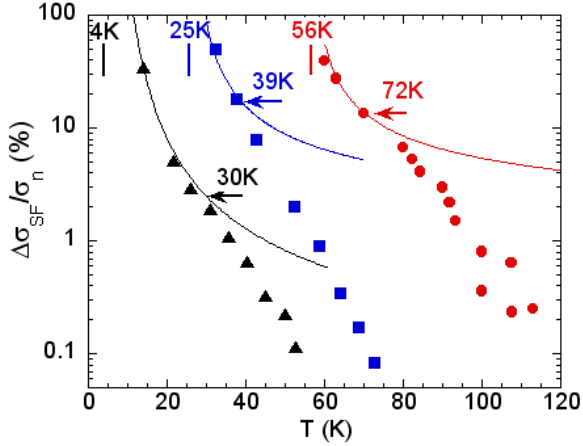}
\caption{(color on line) The SCF conductivity normalized 
to the value in the normal state is plotted versus $T$ for the UD57 samples either 
pure (circles) or irradiated with $T_c$ values decreased down to 26.8 K (squares) and 11K (triangles). 
The vertical bars indicate the estimated values of $T_{KT}$ which are slightly lower than the values of $T_{c}$ (taken at the mid-point of the superconducting transition). The arrows are for the mean-field
temperature $T_{c}^{MF}$ estimated here from the deviation to the fitting curves using Eq.\ref{Eq.formulaKT} 
(full lines).}
\label{Fig.HN-fits}
\end{figure}

We have thus tried to fit the data obtained for the SCF conductivity 
in the pure and irradiated UD57 samples in Fig.\ref{Fig.HN-fits}, where $\Delta\sigma_{SF}/\sigma_{n}$
are plotted versus $T$ in a semi-log scale. Only the very small number of data between $T_{KT}\sim T_c$ 
and the sharp downturn of $\Delta\sigma_{SF}$ are pertinent in such fits. In the pure UD57 sample, within the
foregoing analysis, a natural upper limit for the fit would be $T_{c0}\simeq72$K which could be assimilated to $T_c^{MF}$. We indeed find 
that the three significant data permit to obtain values for $A$ and $b$ for which the fitted 
function deviates from the data above $T_{c0}$. If we take the same criterion to estimate $T_{c}^{MF}$ in 
the other samples, we get the values for $\tau_c$ reported in table \ref{Table_parametersHN}. As expected, 
we find that $\tau_c$ increases with disorder, more than a factor 10 between the pure and the most irradiated sample. 

It is clear that the limited analysis done above is not sufficient to prove \textit{per-se} that the increased magnitude of the SCF can be attributed to phase fluctuations. Both phase fluctuations and amplitude fluctuations could be emphasized altogether, as claimed by some authors \cite{Tallon-2010}. Hopefully with the larger pulse fields which become available now, more data points between $T_{KT}$ and $T_{c}^{MF}$ could become accessible and would allow to better test the applicability of  Eq.\ref{Eq.formulaKT} and to get reliable determinations of the different parameters. The fast suppression of the SCF at high $T$, similar to that found for the optimally doped samples, will be discussed in section VIII.A.

\begin{table}[h]
\caption{Parameters extracted from the fits of the low $T$ data of Fig.\ref{Fig.HN-fits} for the different 
UD57 samples}
\label{Table_parametersHN}
\begin{ruledtabular}
\begin{tabular}{cccc}
samples &  pure  &  irr1  &  irr2 \\
\hline
$T_{KT}$ (K) &  56  &  26.5  &  4\\
$T_{c}^{MF}$ (K) &  72  &  39  &  30\\
$\tau_{c}$  &  0.22  &  0.5  &  2.6\\
$2\alpha^{\prime}$ &  $\simeq 0.5$  &  $\simeq 1$  &  $\simeq 2$\\
\end{tabular}
\end{ruledtabular}
\end{table}

\section{Fluctuation magnetoconductivity} 
It is also interesting to look more carefully on the way the SCF are suppressed 
by the magnetic field. 
Let us recall here that the dependence of the magnetoconductivity with $H$ and $T$ 
has been extensively studied in high-$T_{c}$ cuprates \cite{Lang, Semba, Holm, Bouquet,Sekirnjak, Ando-PRL2002}. This has been often preferred to the study 
of paraconductivity as its value is often considered as weakly 
dependent on the normal state magnetoconductivity. In fact this is not really correct even at low 
magnetic fields since both quantities display inital $H^{2}$ variations as shown in Fig.\ref{Fig.decomposition}.
It is thus necessary to subtract the normal state contribution as expressed in 
Eq.\ref{def-magnetoconductivity} and to consider the difference
\begin{eqnarray}
\Delta\sigma_{H}(T,H) & = & \Delta\sigma(T,H)-\Delta\sigma_{n}(T,H)\, \nonumber \\
& = & \Delta\sigma_{SF}(T,H)-\Delta\sigma_{SF}(T,0)
\label{Eq.correct-magneto}
\end{eqnarray}  

Within the GL theory, the evolution of the fluctuating magnetoconductivity with $H$
comes from the pair-breaking effect which leads to a $T_{c}$ suppression. Different contributions must 
be taken into account to completely explicit the effect of a magnetic field on the excess conductivity 
\cite{Hikami, Sekirnjak, Matsuda, Semba, Holm, Bouquet}. 
Nevertheless it seems legitimate to neglect the Maki-Thompson contribution which already 
does not contribute to 
the paraconductivity in zero field. The Aslamazov-Larkin contribution is the sum of two 
different contributions resulting from interactions of the magnetic field with the carrier orbital (ALO) 
and spin (Zeeman) degrees of freedom. The former vanishes for applied fields in the ab plane, and 
the remaining Zeeman term is usually found much smaller than the ALO term for $H //c$ and low enough magnetic fields \cite{Sekirnjak, Holm}.

In the layered superconductor model the ALO fluctuation magnetoconductivity
can then be written as \cite{Sekirnjak, Holm}:
\begin{eqnarray}
\Delta\sigma_{H}^{ALO}(T,H) = \frac{e^{2}}{8\hslash }\frac{1}{h^{2}} \int_0^{2\pi/s} \epsilon_{k}\bigg[ %
\Psi \left(\frac{1}{2}+\frac{\epsilon_{k}}{2h} \right) \nonumber \\
 - \Psi\left(\frac{\epsilon_{k}}{2h} \right)- \frac{h}{\epsilon_{k}}\bigg] \frac{dk}{2\pi} -%
\Delta\sigma^{LD}
\label{Eq.magnetoALO}
\end{eqnarray}
Here $\epsilon_{k}=\epsilon[1+\alpha(1-cosks)]$, where $\alpha$ is the coupling parameter defined in VI.A
and $k$ is the momentum parallel to the magnetic field H. $\Psi$ is the di-gamma function, and 
$h \equiv H/H_{c2}(0)$.
This expression assumes that the temperature dependence of $H_{c2}(T)$ is simply given by:
\begin{equation} 
H_{c2}(T)= \Phi_{0}/2 \pi \xi(T)^{2} = \epsilon H_{c2}(0)
\label{H_c2}
\end{equation}
This holds as long as the behavior in magnetic field is set by the size of $\xi(T)$. But when the 
magnetic field becomes large enough, the magnetic length $l_B = (\hslash/2eH)^{1/2}$ enters into play 
and overcomes the variation of $\xi(T)$. The crossover between these two regimes occurs for a field
$H^{\ast}$ such as 
\begin{equation}
H^{\star} \simeq \epsilon H_{c2}(0)
\label{Eq.H*}
\end{equation} 
This magnetic field $H^{\ast}(T)$ defined 
above $T_c$ has been called the "ghost critical field" by Kapitulnik \textit{et al} \cite{Kapitulnik} 
as it mirrors the upper critical field defined below $T_{c}$. For $H > H^{\ast}(T)$, the variation with $H$ 
is governed by the magnetic length $l_{B}$ as recently evidenced by Nernst measurements in SC disordered
films \cite{Pourret-PRB2007}.
 
\begin{figure}
\centering
\includegraphics[width=8cm]{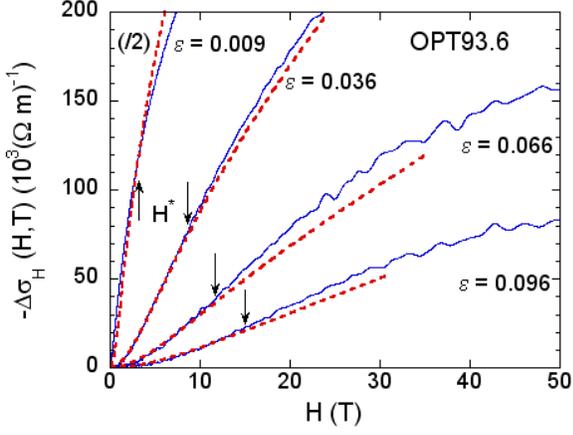}
\caption{(color on line) Evolution of the fluctuation magnetoconductivity $-\Delta\sigma_{H}(T,H) =\Delta\sigma_{SF}(T,0)-\Delta\sigma_{SF}(T,H)$
as a function of $H$ for the OPT93.6 sample at different temperatures, 94.4, 97, 100 and 103.4K. The dotted lines represent the computed results from Eq.\ref{Eq.magnetoALO} with  $H_{c2}(0)=180(10)$T. They deviate from the data beyond the $H^{\star}$ field values shown by arrows and reported in Fig.\ref{Fig_H-ghost}.}
\label{Fig_magnetoALO-OPT}
\end{figure}
In order to analyse the experimental data, we use the procedure described in ref.\cite{Kapitulnik}. We first 
determine the only adjustable parameter $H_{c2}(0)$ by matching the low-field part of the data for all 
values of $\epsilon$. We then introduce the higher field values of $-\Delta\sigma_{H}(T,H)$ computed from 
Eq.\ref{Eq.magnetoALO}, which allows us to define the ghost field  $H^{\star}$ above which a deviation 
from the experimental data occurs.

\begin{figure}[h]
\centering
\includegraphics[width=7cm]{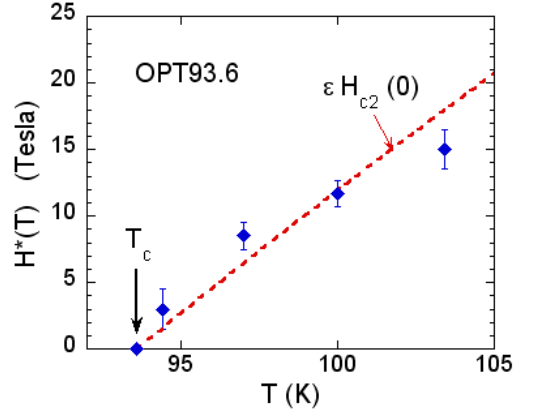}
\caption{(color on line) Values of $H^{\ast}$ at which Eq.\ref{Eq.magnetoALO} deviates from the experimental results. Those 
are compared to the expected linear dependence of Eq.\ref{Eq.H*} using the determined value of $H_{c2}(0)=180(10)$T.}
\label{Fig_H-ghost}
\end{figure}

This is illustrated for the OPT93.6 sample in Fig.\ref{Fig_magnetoALO-OPT} where the evolution of
$-\Delta\sigma_{H}=\Delta\sigma_{SF}(T,0)-\Delta\sigma_{SF}(T,H)$ is plotted versus $H$ at different
temperatures ranging from 94.4 to 103.4K, corresponding to $\epsilon$ values from 0.0085 to $\sim 0.1$. 
In this temperature range, good fits of the low-field data can be achieved with $H_{c2}(0)=180(10)$T. 
One can also see that the agreement deteriorates at larger fields $H^{*}$ with increasing temperatures. 
This is in reasonable agreement with what is expected from Eq.\ref{Eq.H*}, as can be seen in 
Fig.\ref{Fig_H-ghost}. \textit{So, the data follow unambiguously the GL analysis and 
enable us to determine $H_{c2}(0)$ reliably.}

At higher $T$, beyond the GL range deduced from the zero field SCF conductivity data, the low field 
data cannot be matched with Eq.\ref{Eq.magnetoALO} with the same value of $H_{c2}(0)$.
One may artificially find a better agreement by increasing $H_{c2}(0)$ with increasing temperature, but 
this is meaningless and only confirms that Eq.\ref{Eq.magnetoALO} is no longer valid beyond the GL region. 
So the choice of temperature range used to fit the data can severely affect the deduced value of 
$H_{c2}(0)$. This might explain why previous attempts to deduce $H_{c2}(0)$ from magnetoconductivity data
give contrasting results for optimal doping (those of ref.\cite{Sekirnjak} are in better agreement with 
ours than those of ref.\cite{Ando-PRL2002}).

\begin{figure}
\centering
\includegraphics[width=8cm]{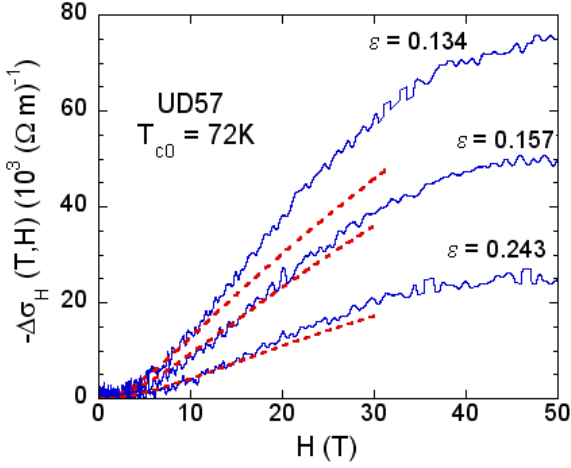}
\caption{(color on line) Evolution of the fluctuation magnetoconductivity $-\Delta\sigma_{H}(T,H) =\Delta\sigma_{SF}(T,0)-\Delta\sigma_{SF}(T,H)$
as a function of $H$ for the UD57 sample at different temperatures, 82.4, 84.2 and 91.8K. The dotted lines represent the computed results using Eq.\ref{Eq.magnetoALO} with $T_{c0}=72$K and $H_{c2}(0)=90\pm10$T.}
\label{Fig_magnetoALO-UD}
\end{figure}

We could repeat the same procedure for all the dopings studied here, which leads to values of 
$H_{c2}(0)$ indicated in table \ref{Table_Hc2-H'c}. In the case of the UD57 sample, the analysis has 
been performed by assuming that the mean-field temperature is 72K, as indicated above.
As seen in Fig.\ref{Fig_magnetoALO-UD}, it is still possible to fit the low-field data reasonably well using 
Eq.\ref{Eq.magnetoALO} in a $T$ range which is found to slightly exceed the GL regime. 
  
We notice that the $H_{c2}(0)$ value determined in this way are surprisingly close to those obtained 
for $H_{c}^{\prime}(0)$ in a totally different way in section IV.B, which are reported as well in 
Table \ref{Table_Hc2-H'c}.
\begin{table}[h]
\caption{Values of $H_{c2}(0)$ extracted from the fluctuation magnetoconductivity. 
They are very close to the values of $H_{c}^{\prime}(0)$ determined in section IV-2. 
The values of $\xi(0)$ are calculated from $H_{c2}(0)$ using Eq.\ref{Eq.H_c2(0)}. 
The respective values of the superconducting gap $\Delta_{SC}$ are estimated from Eq.\ref{Eq.xi-Delta}.}
\label{Table_Hc2-H'c}
\begin{ruledtabular}
\begin{tabular}{ccccc}
samples &  UD57  &  UD85  &  OPT93.6 & OD92.5 \\
\hline
$H_{c2}(0)$ (T) &  90(10)  &  125(5)  &  180(10) & 200(10) \\
$H_{c}^{\prime}(0)$ (T) &  86(10)  &  115(15)  &  155(10) & 207(10) \\
$\xi(0)$(nm) & 1.9 & 1.60 & 1.33 & 1.26 \\
$2\Delta_{SC}/\beta$ (meV) & 46 & 50 & 66 & 69.5 \\
\end{tabular}
\end{ruledtabular}
\end{table}
\textit{The very important result of this analysis is to unambiguously show that $H_{c2}(0)$ increases and thus $\xi(0)$ decreases with increasing doping in YBCO}. 

The coherence length is related to the superconducting gap $\Delta_{SC}$ through:
\begin{equation}
\xi(0) = \beta \left(\frac{\hbar v_F}{\pi \Delta_{SC}}\right)
\label{Eq.xi-Delta}
\end{equation}
with $\beta=1$ for s-wave superconductors. By assuming that the Fermi velocity $v_F$ is weakly dependent on doping as found 
in different cuprates \cite{Zhou} and equal to \cite{Fournier} $v_F \simeq 2.2\ 10^5$m.s$^{-1}$, 
we obtain the values of $2\Delta_{SC}/\beta$ indicated in Table \ref{Table_Hc2-H'c}. Independently of the 
precise value of $\beta$, our results demonstrate 
that the superconducting gap is closely related to $T_{c}$ in the doping range
$\sim 0.09$ to $\sim 0.17$ considered here. 

Recent STM measurements have shown that 
$\Delta_{SC} \simeq 20$meV for an overdoped Bi2212 sample with $T_c=63$K \cite{Yazdani}, yielding a 
ratio $2\Delta_{SC}/k_BT_c = 7.4$. A similar ratio is also found for the "small" gap $\Delta=6.7\pm1.6$meV
identified in the STM spectra of an underdoped Bi2201 sample with $T_c=15$K \cite{Hudson}.
We notice that for $\beta=1$ in Eq.\ref{Eq.xi-Delta}, the data of Table \ref{Table_Hc2-H'c} would 
also correspond to a similar gap magnitude $2\Delta_{SC}\simeq 8k_BT_c$ whatever the doping.
This is a strong indication that \textit{the gap determined here
can thus be assimilated to the "small" gap detected recently by different techniques} \cite{Hufner}.

\section{Suppression of SC fluctuations by temperature or magnetic field}
We shall consider now more specifically the sharp decrease of SCF found versus temperature in 
Fig.\ref{Fig.LD-pure} and the onset of SCF at $T_c^{\prime}$ and $H_c^{\prime}$.
\subsection{Temperature}
\begin{figure}
\centering
\includegraphics[width=8cm]{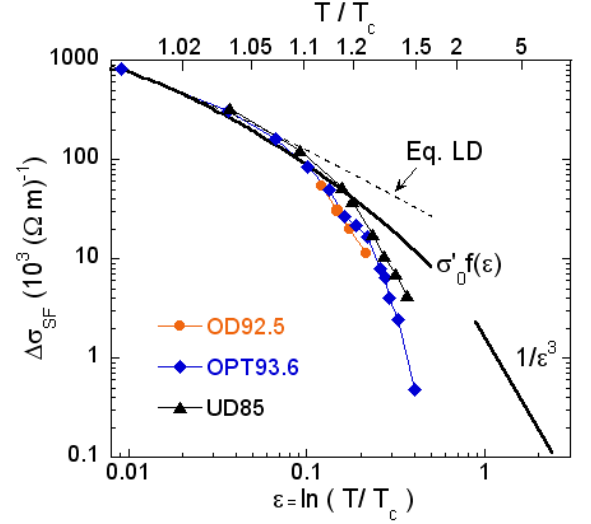}
\caption{(color on line) Comparison of the temperature dependence of $\Delta\sigma_{SF}$ for the 
OD92.5, OPT93.6 and UD85 samples with Eq.\ref{Eq.AL-generalized} (full line) that takes into account short-wavelength cut-off in the fluctuation spectrum \cite{Reggiani, Glatz-2010}.  The dashed line is for the Lawrence-Doniach expression (Eq.\ref{Eq.LD}).}
\label{Fig.AL-cutoff}
\end{figure}

Within the Ginzburg-Landau description, there is a priori no upper temperature limit for the existence 
of fluctuations, that are expected to survive far above $T_{c}$ in the normal state. 
However it has been pointed out very early that a rapid attenuation of the fluctuations may occur 
for $T \gg T_{c}$ in short coherence-length systems, as the GL theory can be put into question 
when $\xi(T)$ becomes comparable to the zero-temperature in-plane coherence
length $\xi(0)$ \cite{Freitas, Hopfengartner, Gauzi}. It has been first argued that a short-wavelength 
cut-off should be taken into account in the fluctuation spectrum. An extension of the AL theory in 
the two dimensional case taking this cut-off into account \cite{Reggiani} gives 
a $T$ dependence of the superconducting fluctuation conductivity as:
\begin{equation}
\Delta\sigma_{SF}(T)=\frac{e^{2}}{16\hslash s} f(\epsilon)= \sigma_0 f(\epsilon)
\label{Eq.AL-generalized}
\end{equation}
where the function $f(\epsilon)$ matches the $1/\epsilon$ first order expression up to $\epsilon \simeq 0.18$
but deviates then to reach the asymptotic limit \cite{Glatz-2010} $f(\epsilon) \propto 1/\epsilon^{3}$. 
Our data can be fitted using the function $f(\epsilon)$ up to $\epsilon \simeq 0.15$. But as shown in 
Fig.\ref{Fig.AL-cutoff}, it then drops much faster well below the expected $\epsilon^{-3}$ dependence,
above temperatures corresponding to coherence lengths $\xi(T) \lesssim 3\xi(0)$. 

Other authors have proposed that a "total-energy" cut-off \cite{Carballeira, Vidal} should be more
appropriate to describe the evolution of the paraconductivity in the high $\epsilon$ region. 
They assigned its origin to the intrinsic constraint that SCF cannot survive when the coherence
length $\xi(T)$ becomes comparable to the superconducting coherence length $\xi_0$. They proposed then to mimic this effect using a phenomenological expression which can be transformed into
\begin{equation}
\Delta\sigma_{SF}(T) \simeq \Delta\sigma^{LD}\left(1-\frac{\epsilon \sqrt{1+2\alpha}}{\epsilon^{C}}\right)^{2}
\label{Eq.energy-cutoff}
\end{equation}
where $\epsilon^{C}=\ln (T^{C}/T_{c})$. Our data in YBCO unambiguously allowed us to demonstrate 
that SCF are heavily, if not totally, suppressed at high $T$, which allowed us to define a 
temperature $T_{c}^{\prime}$ above which their detection becomes nearly impossible. 
This $T_{c}^{\prime}$ value could in principle be obtained as well using the phenomenological 
Eq.\ref{Eq.energy-cutoff}. However this equation privileges the cut-off behaviour and does not reproduce 
the low $T$ regime where we have established the validity of the Ginzburg-Landau approach. 

\begin{figure}
\centering
\includegraphics[width=8cm]{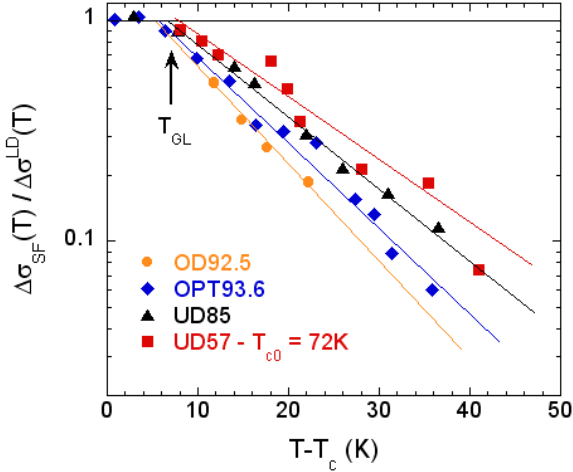}
\caption{(color on line) The ratio $\Delta\sigma_{SF}(T)/\Delta\sigma_{LD}(T)$ is plotted versus $T-T_{c}$ 
in a semi-logarithmic scale for all the pure samples studied. $\Delta\sigma_{LD}(T)$ is calculated from 
Eq.\ref{Eq.LD} using the value of $\xi_c=0.89$\AA\,determined in section VI-A. $T_{GL}$ indicates 
the upper bound for the GL regime. In the case of the most underdoped UD57 sample, we have assumed 
that the phase fluctuation regime disappears at $T_{c0}=72$K (see section VI). The full lines are exponential fits of the data above $T_{GL}$.}
\label{Fig.deviationsLD}
\end{figure}

In order to analyse more precisely the attenuation of the superconducting fluctuations with temperature,
we have therefore plotted in Fig.\ref{Fig.deviationsLD} the values of $\Delta\sigma_{SF}(T)$ normalized to 
those expected from the LD formula versus $T-T_c$ in a semilogarithmic scale. For all the samples,
we observe that the SCF conductivity vanishes exponentially for $T>T_{GL}$, as
\begin{equation}
\Delta\sigma_{SF}(T) = \Delta\sigma^{LD}(T) \exp \left(-\frac{T-T_{GL}}{T_{0}}\right)
\label{Eq.sigma-T}
\end{equation} 
Here, $T_{GL}$, the upper temperature for the Ginzburg-Landau regime  exceeds $T_{c}$ of about 5 to 7K. 
Let us recall here that, for the most underdoped sample, we have replaced $T_c$ by $T_{c0}=72$K, the mean field
temperature determined above. The exponential decay rate $T_0$ of the SCF increases from 10 to 15K 
with decreasing doping. 

In our previous report \cite{RA-HF}, we had proposed that the total suppression of SCF in the 
cuprates could be associated with an intrinsic ultimate possibility to thermally induce pairs in 
these systems. The simultaneous analysis of the Nernst, paraconductivity and magnetoconductivity
data allow us to underline that $\xi(0)$ is intrinsically 
very small in these compounds and of the same order as the mean distance $d$ between the hole carriers, 
that is about 10\AA\,  for instance in optimally doped cuprates with a hole content of 0.16 per Cu. 
Therefore pairing with shorter coherence length scale would correspond to a thermal excitation of 
isolated Bose pairs. The total suppression of paraconductivity and Nernst 
signals above $T_{c}^{\prime}$ therefore means that the density of such excited pairs drops to zero 
already at $T_{c}^{\prime}$. This fits with the idea that the energy of fluctuations which can be 
thermally excited is bounded with a sharp cut-off at an energy $k_{B}T_{c}^{\prime}$. 

In usual BCS materials, for which the density of carriers per atom is of the order of unity, 
$\xi$ is large enough with respect to the atomic distance $a$, so that SCF fluctuations could 
be thermally excited as long as $\xi(T) \gg a$, which agrees with the observation that SCF survive up to 
at least $T \simeq 30T_{c}$ in amorphous superconducting films \cite{Pourret-NPhys2006}.

The existence of such a sharp cut-off at $T_{c}^{\prime}$ is not restricted to pure samples but also applies for disordered ones. This similarity appears very clearly in Fig.\ref{Fig.HN-fits} for the
underdoped irradiated samples where the excess conductivity above $T_{c0}$ nearly parallel that of the 
pure sample and vanish exponentially with the same decay rate.

It is quite intriguing to see this quasi "universality" of the SCF attenuation in 
a $T$ range where one would rather expect a response highly sensitive to the
peculiar features of the systems of interest. This exponential decay means that a $T$ range of 
the order of $5T_{0}$ is universally required to suppress the fluctuations above $T_{c}$, or $T_{c0}$ 
in the underdoped cases.   
Independently of the physical meaning of this representation, this shows that the SCF
vanish similarly with increasing $T$ for all the hole contents. This confirms that the 
pseudogap state has \textit{no specific incidence on the range of SCF}. 
This observation contradicts the recent
proposition \cite{Levchenko-2010} which attributes the rapid suppression of superconducting 
fluctuations evidenced by Nernst effect and conductivity measurements in underdoped LaSrCuO 
\cite{Wang-PRB2006, Leridon-PRB2007} to the presence of the pseudogap. All the detailed analysis of 
the data proposed here rather substantiates the conclusion done previously from the simple comparison 
of the $T_{c}^{\prime}$ and $T^{\star}$ lines \cite{Alloul-EPL2010} that \textit{these two lines are underlining 
two independent phenomena in the phase diagram of cuprates}.

\subsection{Magnetic fields}
In the analysis of the fluctuation magnetoconductivity done above, we have been able to fit the data 
using Eq.\ref{Eq.magnetoALO} as long as $H \lesssim H^{\ast}$ for which the Ginzburg-Landau coherence 
length $\xi(T)$ becomes comparable to the magnetic length $l_{B}$.

\begin{figure}[t]
\centering
\includegraphics[width=8cm]{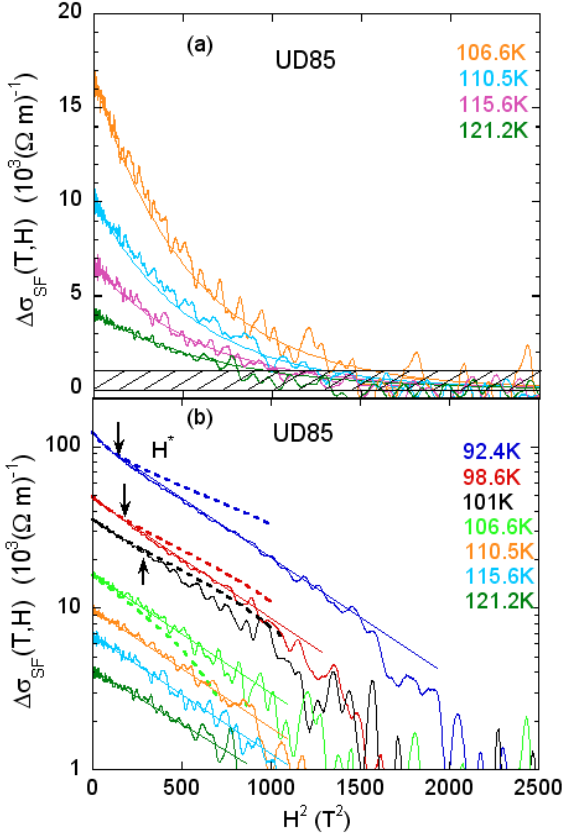}
\caption{(color on line) SC fluctuation contribution to the conductivity $\Delta\sigma_{SF}(T,H)$
plotted versus $H^{2}$ for the UD85 sample. In (a), the data plotted in a linear scale for the high-$T$
regime can be fitted by the exponential relationship of Eq.\ref{Eq.magneto-H^2} with $H_0 \simeq 25$T. 
In (b) the data are plotted in a semi-logarithmic scale for all the temperatures investigated. The 
dotted lines are curves using Eq.\ref{Eq.magnetoALO} with $H_{c2}(0)=125(5)$T which deviate from 
the experimental data for $H>H^{\star}$ pointed by arrows. The straight lines are exponential fits 
of the high field data with $H_0 = 25 \pm3$ Tesla whatever $T$.}
\label{Fig.sigma-H^2-UD85-lin-log}
\end{figure}
Well beyond the Ginzburg-Landau regime, 
for $\epsilon \gtrsim 0.2$, where the coherence length is strongly reduced by temperature, we expect 
that the fluctuation magnetoconductivity cannot be described any longer by Eq.\ref{Eq.magnetoALO}. Some data
taken in this regime are illustrated in the case of the UD85 sample in 
Fig.\ref{Fig.sigma-H^2-UD85-lin-log}(a).

There it can be seen that the excess conductivity appears 
to decay exponentially with the magnetic field as
\begin{equation}
\Delta \sigma_{SF}(T,H) = \Delta \sigma_{SF}(T,0)\ \exp [-(H/H_{0})^{2}] 
\label{Eq.magneto-H^2}
\end{equation}
This sharp exponential decay confirms that $H_{c}^{\prime}(T)$ can indeed be reliably defined and is not 
so dependent on the criterion used (we defined it here and in ref.\cite{Alloul-EPL2010} for $\Delta\sigma_{SF}=1\,10^3 (\Omega.m)^{-1}$)
 
\begin{figure}
\centering
\includegraphics[width=8cm]{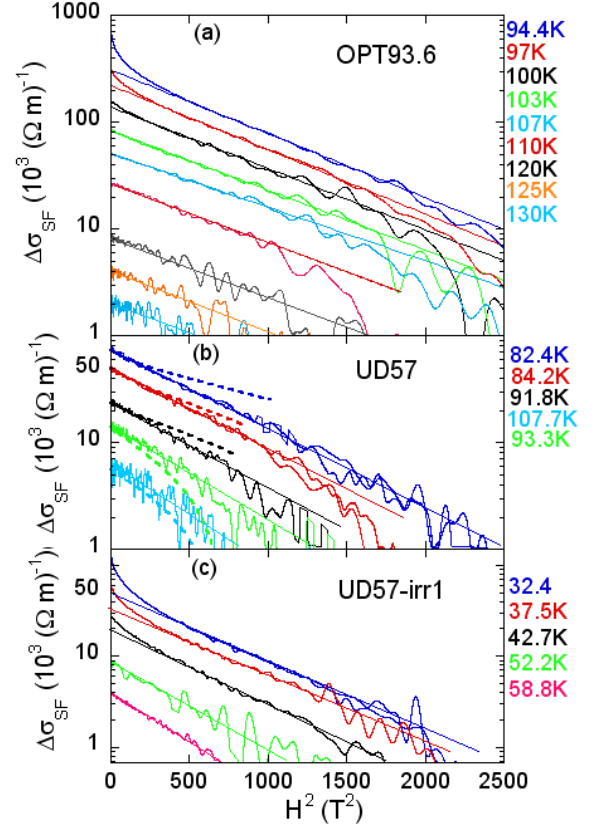}
\caption{(color on line) SC fluctuation contribution to the conductivity $\Delta\sigma_{SF}(T,H)$
plotted versus $H^{2}$ in a semilog  scale for (a) the OPT93.6 sample, (b) the UD57 sample and (c) the 
UD57 irradiated sample with $T_c=25$K.  The full lines are exponential fits according to 
Eq.\ref{Eq.magneto-H^2} which do not take into account the low field data at low 
$T$. For all these samples, we find $H_0 = 25 \pm5$Tesla at all temperatures. For 
the pure UD57 sample, we have also indicated the matching curves taken from Eq.\ref{Eq.magnetoALO} with 
$H_{c2}(0)=90(10)$T in order to better visualize the deviations at larger fields. Here again it is seen that 
Eq.\ref{Eq.magnetoALO} does not fit the data for $T \gtrsim 93.3$K, even at low fields, if one keeps the 
same value of $H_{c2}(0)$ (see discussion in section VII).}
\label{Fig_logsigma-H^2}
\end{figure}

In order to better visualize how the SCF are suppressed by magnetic fields in the whole $T$ 
range, we have then plotted $\Delta\sigma_{SF}$ versus $H^2$ in a semilogarithmic scale in 
Fig.\ref{Fig.sigma-H^2-UD85-lin-log}(b) for the UD85 sample. For the lowest temperatures, we find that
Eq.\ref{Eq.magnetoALO} applies with $H_{c2}=125(5)$T as long as $H<H^{\star}$.  
It is intriguing to see on this plot that the decay of $\Delta\sigma_{SF}$ evolves then smoothly towards 
an exponential behaviour with nearly the same value of $H_0$ as found at higher temperatures.
The same type of evolution is observed for all the samples, pure or irradiated.  
$H_0$ remains nearly constant whatever the temperature, doping 
or disorder level with $H_{0}=25 \pm 5$T as can be seen in Fig.\ref{Fig_logsigma-H^2}.

To our knowledge, such an exponential suppression of the magnetoconductivity has never been reported 
experimentally nor predicted theoretically. 

\section{Discussion and conclusions}
We have done here a set of measurements where the  normal state
magnetoresistance of YBa$_2$Cu$_3$O$_{6+x}$ could be followed down in temperature from
the high $T$ totally non superconducting state. This allowed us to monitor the progressive advent 
of fluctuation contributions to the conductivity above $T_{c}$. We could not therefore study 
the close vicinity of $T_{c}$, that is the 
3D critical exponents. However this experiment quite uniquely allowed us to study
the variation of SCF from the 3D to the 2D higher $T$ regime. We could
evidence that the Ginzburg Landau regime applies near $T_{c}$ for
optimally doped samples, while for underdoped ones phase fluctuations might play
a role in a narrow $T$ range above $T_{c}$. Above those $T$
ranges the SCF are highly damped, which reveals the intrinsic microscopic
limitations of the pairing at high temperatures. We have also evidenced that
disorder increases the phase fluctuation regime above $T_{c}$. We shall  
summarize below the most important conclusions and questions which
arise from this work. 

\subparagraph{Normal state properties in the pseudogap phase}
In section III we definitely evidenced that a 60 tesla field is not sufficient to suppress totally the 3D
superconductivity at $T_{c}$ in the pure 123 phases, even for underdoped samples with $T_{c}\simeq 60$K, 
so that the normal state transport properties are only accessible above. The SCF could only be 
suppressed fully with 60 tesla in the presence of strong disorder reducing $T_{c}$ down to $\sim 4$K. 

In the pure UD57 sample we had demonstrated that the resistivity keeps a metallic behaviour at low 
$T$ in large applied fields \cite{RA-HF}. This hole content is slightly lower than that on which 
maximal quantum oscillations have been observed at low $T$ and high applied fields \cite{Doiron-Leyraud}.
From the negative 
Hall effect detected in these experiments \cite{Leboeuf}, a reconstruction of the Fermi surface with the
appearance of an electron pocket has been proposed.  Here we evidenced that the simple relation between 
the magnetoresistance and the Hall effect which had been established above $ \sim130$K in the past
\cite{Harris} has a validity which extends nearly down to $T_{c}$, without any singular behaviour both for 
this YBCO$_{6.6}$ composition and for an optimally doped sample. This is in rather good agreement 
with the fact that, for underdoped $T_c=57$K samples, the high field Hall constant becomes negative 
only below the zero field $T_{c}$, and that the Fermi surface reconstruction only arises deep into 
the SC state, in fields which are however insufficient to totally suppress the SCF.

\subparagraph{Ginzburg-Landau regime: critical fields and gaps }
For samples around optimal doping, the quantitative comparative analysis of
the measured SCF contribution to the zero field conductivity and of the
off-diagonal Peltier term $\alpha_{xy}$ has been found in total agreement with the GL approach
for 2D Gaussian order parameter fluctuations (section VI.B). The data perfectly fit the
leading order 2D Aslamazov Larkin contribution up to $T\simeq 1.1T_{c}$
using the $c$ lattice constant as the mean spacing between the CuO$_{2}$ bilayers. It can be fitted as
well up to $\sim1.2 T_{c}$ if higher order corrections are taken into account.
This analysis allows us to deduce values of $\xi(0)$ and of 
$H_{c2}(0)$ versus doping. 

The analysis of the fluctuation magnetoconductivity in
this GL regime allows us to determine $H_{c2}(0)$ independently in section VII. 
The good agreement between these different values establishes the perfect consistency of our data analyses. 
A very important result obtained here is that the deduced \textit{%
superconducting gap increases smoothly with increasing hole doping from the
underdoped  to the overdoped regime}.

Let us recall that energy resolved spectroscopies have evidenced spectral
gaps in the SC state which increase with decreasing doping while here we
find a gap which rather follows the same trend as $T_{c}$. For overdoped
samples the local density of states (LDOS) has coherence peaks \cite{Renner, Yazdani}
and exhibits the $\mathbf{k}$ dependence expected for $d$ wave pairing, which distinguishes the nodal and
antinodal regions \cite{review-ARPES}. Above $T_{c}$ a small dip in the LDOS remains and has
been assigned to the pseudogap, but should be attributed to SCF, as we have shown
that in this range the pseudogap disappears \cite{Alloul-EPL2010}.

But in the underdoped cases a large gap is found to persist then well
above $T_{c}$, while at low energies the LDOS becomes nearly independent of
local disorder \cite{McElroy, Hudson}. A large debate has been raging recently as various spectroscopy
data have suggested that a smaller gap exists, visualized in the
nodal regions by Raman spectroscopy \cite{LeTacon} or obtained by discriminating different spectral
weights in the ARPES or STM spectra \cite{Kondo, Hudson}. Our deduction here, that an important SC 
property deduced from thermodynamic considerations, that is the critical field  $H_{c2}(0)$, is governed 
by a gap which follows $T_{c}$, reinforces then the idea
that \textit{the pseudogap is connected with the large gap detected by STM and ARPES
on cuprate sample surfaces in the underdoped regions of their phase diagram}. Conversely it can be seen that the gap 
magnitudes deduced from our data scale quite nicely with the smaller gaps obtained by STM 
\cite{Yazdani, Hudson}.

\subparagraph{Phase coherence and phase fluctuations}
For the $T_{c}=57$K underdoped sample, well into the pseudogap phase, the SCF
paraconductivity and Nernst coefficient are found in section VI.C both much larger than
expected for Gaussian fluctuations in a range of temperatures of the order of 15K above $T_{c}$, 
which points for the occurrence of phase fluctuations .
This range increases markedly if disorder is used to decrease
$T_{c}$ and can become as large as 40K when $T_{c}$ has been
depressed down to $T_{c}=5$K. These results are therefore consistent with the proposal
done by Emery and Kivelson \cite{Emery}, that in the underdoped regime, controlled disorder reduces 
the phase coherence. The regime where phase fluctuations might play an important role occurs 
then between the 3D $T_{c}$ up to a mean field temperature which we can assimilate to $T_{c0}$. In this limited $T$ range we do not 
have sufficiently accurate measurements, nor theoretically established firm criteria 
to go beyond qualitative observations. 

We noticed however that the Nernst signal is more enhanced than the
excess fluctuation conductivity with respect to expectations for Gaussian fluctuations.
More work, both theoretical and experimental, is required to decide the
possible importance of vortex contributions to the Nernst effect in this
phase fluctuation regime and/or other possibilities such as the enhancement of SCF by AF spin fluctuations \cite{Kontani}. However this enhancement of Nernst effect with
respect to excess conductivity decreases for $T>T_{c0}$. So while it has
been recently proposed that Nernst measurements were among the best approaches to
probe the extension of the SCF above $T_{c}$ \cite{Wang-PRB2006}, we demonstrated here that
those are indeed not as powerful as expected initially for pure YBCO as
they are limited by the need of an independent determination of the normal
state Nernst coefficient. The latter is not as small as could be anticipated
from the Sondheimer cancellation rule which applies only for classical
metals \cite{Behnia-Nernst, Vojta}. For the conductivity measurements, our approach using high
fields and the former knowledge of the high $T$ magnetoresistance permitted
to circumvent the corresponding difficulties, so that the SCF could be
followed until they are fully suppressed at high $T$. 

\subparagraph{Suppression of SCF at high $T$ and pairing energies}
We have evidenced that in all samples, pure or disordered, and for all
dopings, the SCF sharply decay with increasing $T$ or $H$, in the ranges
where SC gaussian fluctuations are dominant. In section IV.B, we could then deduce for all the
samples a curve $H_{c}^{\prime }(T)$ ending at $H_{c}^{\prime}(T_{c}^{\prime})=0$, 
which delineates the $(H,T)$ plane region beyond
which SCF are totally suppressed. 

In section VI.A, the SCF are found to be much more rapidly depressed
than in thin films of classical s-wave metallic superconductors for which
SCF are detected even for $T\gg T_{c}$ \cite{Pourret-NPhys2006}. This provides a strong support 
for our preliminary suggestion \cite{RA-HF} that the $H_{c}^{\prime}(T)$ curve delineates the regime
where microscopic considerations specific to the cuprate physics prohibit SC
pairing. We propose in section VIII.A that the spatial pair extension at high $T$ is
limited by the actual density of carriers available for pairing, \textit{so
that a lower bound of $\xi(T)$ could be linked with the distance
between doped holes.}

The suppression of the fluctuation conductivity is found in section VIII to display a phenomenological
exponential decay in $\exp[-(T_{c}-T_{GL})/T_{0}]$ and $\exp(-(H/H_{0})^{2})$,
with values of $T_{0}$ and $H_{0}$ which are not markedly dependent on the
doping and disorder. This suggests as well that there is a sharp energy
cut-off at $k_{B}T_{c}^{\prime }$ which is shifted by the magnetic energy
increase which scales with $H_{0}^{2}$. The energy balance is such that SC
pairs cannot be thermally excited any more above $T_{c}^{\prime }$ or $H_{c}^{\prime}(T)$.
Let us notice as well that the extrapolated values of $H_{c}^{\prime}(0)$ have
been found to be nearly identical to those obtained for $H_{c2}(0)$, which gives confirmation 
that \textit{both are connected with the pairing energy}. All these consistent
deductions give weight to the present analysis.

\subparagraph{Influence of disorder and generic PD of cuprates}
It has been found by STM that cuprates (or at least Bi2212 surfaces)
displayed a short range disorder, visible for instance as a spatial
distribution of spectral gaps \cite{Pan}. These observations have been
questioned as being non generic, as NMR data indicate that YBCO is not as disordered \cite{Bobroff-PRL2002}, 
hence the metallic behaviour observed for YBCO$_{6.6}$ \cite{RA-HF}. This has justified our use of YBCO to 
study the pure cuprate behaviour and the incidence of disorder \cite{FRA-PRL2001}. We have as well
shown that controlled disorder affects drastically the transport properties.
Indeed similar upturns of $\rho(T)$ have been found for controlled disorder
in YBCO and in some pure cuprate families, which indicated the occurence of 
intrinsic disorder in those families \cite{FRA-EPL-MIT}. 

The influence of such disorder on SCF has been thoroughly studied in section IV.C 
and we have shown that the local pair formation underlined by the 
$T_{c}^{\prime}$ line is only moderately affected, while the bulk $T_{c}$,
that is the SC pair coherence can be severely reduced by disorder.
Our results allow us to draw important conclusions on the cuprate phase
diagram which we had specifically emphasized in a preliminary report 
\cite{Alloul-EPL2010}. We could determine that the pseudogap line crosses the 
$T_{c}^{\prime}$ line at optimal doping, which establishes unambiguously
that the pseudogap is not the onset of pairing. The results presented here
re-enforce completely this conclusion as \textit{the fluctuations are
similarly limited in field and temperature independently of the pseudogap},
though they are enhanced in magnitude in the underdoped regime.

We want to insist here that specific effects induced by disorder are
certainly at the origin of many confusions in the study of HTSC. It is
interesting to mention here very recent STM data taken on classical metallic
films in presence of large disorder \cite{Sacepe1}. LDOS measurements reveal strong spatial 
inhomogeneities of the superconducting gap. The remarkable finding
is that the gap magnitude is not much affected when increasing $T$ through $T_c$
while the coherence peaks in the one-particle LDOS disappear. While pairs should be thermally 
excited and fluctuating above $T_{c}$, those appear to be localized by disorder as preformed pairs.
The authors call "pseudogap" the reduction of LDOS detected above $T_c$ \cite{Sacepe2}. This  gap, which is induced by superconducting fluctuations and 
favored by the vicinity of the superconductor insulator transition in the most disordered samples, 
has no relation with the situation encountered 
in clean HTSC, for which the pseudogap is not due to SC pairing and has no connection
whatsoever with disorder \cite{RMP}. This experiment is however quite striking as
it demonstrates how disorder can produce phenomena which can be easily
confused with the pseudogap which characterizes the properties of clean 
cuprates.

\begin{figure}[t]
\centering
\includegraphics[width=7cm]{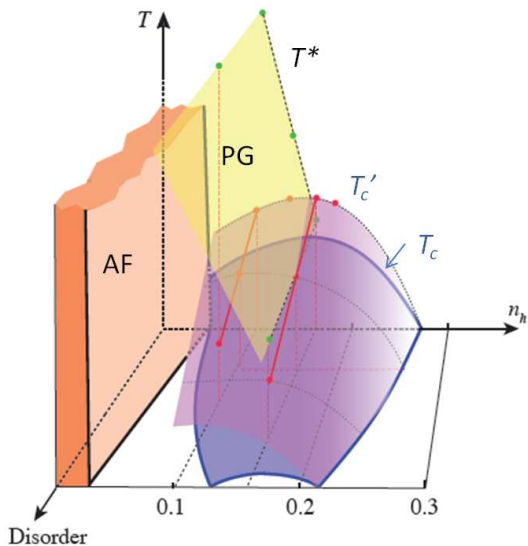}
\caption{(color on line) Phase diagram constructed on the data points
obtained here, showing the evolution of $T_{c}^{\prime}$ the onset of
SCF, with doping and disorder. The fact that the pseudogap and the SCF
surfaces intersect each other near optimum doping in the clean limit is
apparent. These surfaces have been limited to experimental ranges where
they have been determined experimentally. In the overdoped regime, data taken on Tl 2201 
indicates that disorder suppresses SC without any anomalous extension of the SCF \cite{FRA-PRL2001}.}
\label{Fig.phase-diagram}
\end{figure}
 
This reinforces our insistence that the cuprate phase diagram has to take
into account the presence of disorder, which we have suggested for long to
explain the anomalously low optimum $T_{c}$ value in some cuprate families.
This 3D phase diagram that we anticipated from previous results probing the
metal insulator transition \cite{FRA-EPL-MIT} and from the recent comparison
of  $T_{c}$ and $T^{\ast }$ is displayed in Fig.\ref{Fig.phase-diagram}. There, in the pure high 
$T_{c}$ systems the occurence of SCF and the difficulty to separate the SC
gap from the pseudogap in zero field experiments justifies that the $T_{c}$
line could often be mistaken as a continuation of the  $T^{\ast}$ line. It
can be also seen there that the respective evolutions with disorder of the
SC dome and of the amplitude of the SCF range explains as well the phase
diagram often shown  in a low $T_{c}$ cuprate family such as Bi2201 \cite{Fisher}. There 
both $T^{\ast}$ and $T_c^{\prime}$ might appear well above the shrunk SC dome probably because
the actual concentration of carriers is not determined independently but
just mapped from the shape of the SC dome \cite{Tallon}. Finally for intermediate
disorder, the enhanced fluctuation regime with respect to $T_{c}$
illustrated in the initial Nernst measurements performed in the La$_{2-x}$Sr$_{x}$CuO$_4$
family can  be reproduced as well \cite{Wang-PRB2006}.

\subparagraph{Conclusion}
In the present work we have performed a thorough quantitative study of the SCF which establishes that 
such data give important determinations of some thermodynamic properties of the SC state of 
high-$T_{c}$ cuprates. Those are not accessible otherwise, as flux flow dominates near $T_{c}$ in the 
vortex liquid phase and the highest fields available so far are not sufficient to reach the normal state 
at $T=0$. This is an illustration that the studies of SCF permit a "fluctuoscopy" \cite{Glatz-2010} 
of the SC state. It has allowed us to demonstrate that the pairing energy and SC gap both increase 
with doping, confirming then that the pseudogap is to be assigned to an independent magnetic order 
or crossover due to the magnetic correlations.

Further experimental work in even higher fields should help to better characterize the regime where disorder governs the SCF and to decide about the possible relevance of phase fluctuations. The quasi universal behaviours found for the suppression of SCF both in temperature and magnetic fields beyond the Ginzburg-Landau regime suggests that pairing is prohibited above an energy scale which is directly linked with microscopic parameters responsible for SC in the cuprates. 

Theoretical works within the various scenarios proposed to explain HTSC are highly desirable to connect our data with the microscopic parameters which govern the pairing energy. Such an approach might be helpful to discriminate between theories. 

\begin{acknowledgments}
We acknowledge D. Colson and A. Forget for providing the single crystals used in this work.
We thank C. Proust and B. Vignolle for their help with the high field pulsed experiments. 
We also thank L. Benfatto for helpful discussions about phase fluctuations and Kosterlitz-Thouless regime, 
and K. Benhia for discussions about the Gaussian fluctuations in thin films. 
This work has been performed within the "Triangle de la Physique" and was supported by 
ANR grant "OXYFONDA" NT05-4 41913. The experiments at LNCMI-Toulouse were funded by the FP7 I3 EuroMagNET. 
\end{acknowledgments}


\begin{thebibliography}{99}
\bibitem{Alloul89} H. Alloul, T. Ohno and P. Mendels, Phys. Rev. Lett. \textbf{63}, 1700 (1989).

\bibitem{Alloul-PRL91} H. Alloul, P. Mendels, H. Casalta, J.F. Marucco, and J. Arabski, Phys. Rev. Lett. \textbf{67}, 3140 (1991).

\bibitem{Renner} Ch. Renner, B. Revaz, J-Y. Genoud, K. Kadowaki and O. Fisher, Phys. Rev. Lett. \textbf{80}, 149 (1998).

\bibitem{review-ARPES} A. Damascelli, Z. Hussain, Z.X. Shen, Rev. Mod. Phys. \textbf{75}, 473 (2003). 

\bibitem{Capan} C. Capan et al., Phys. Rev. Lett. \textbf{88}, 056601 (2002).

\bibitem{Wang-PRB2006} Y. Wang, L. Li and N.P. Ong, Phys. Rev. B \textbf{73}, 024510 (2006).

\bibitem{FRA-Nernst} F.Rullier-Albenque, R. Tourbot, H. Alloul, P. Lejay, D. Colson, and A. Forget,Phys. Rev. Lett. \textbf{96}, 067002 (2006).

\bibitem{Corson} J. Corson, R. Mallozzi, J. Orenstein, J. N. Eckstein, I. Bozovic, Nature \textbf{398}, 221 (1999).

\bibitem{Emery} V.J. Emery and S.A Kivelson, Nature \textbf{374}, 434 (1995). 

\bibitem{Wang-PRL2} Yayu Wang, Lu Li, M. J. Naughton, G. D. Gu, S. Uchida, and N. P. Ong, Phys. Rev. Lett. \textbf{95}, 247002 (2005).

\bibitem{Li-PRB2010} L. Li, Y. Wang, S. Komiya, S. Ono, Y. Ando, G.D. Gu and N.P. Ong, Phys. Rev. B
\textbf{81}, 054510 (2010).

\bibitem{stripes} V. J. Emery, S. A. Kivelson, and J. M. Tranquada, PNAS \textbf{96}, 8814 (1999).

\bibitem{STM-nematic}T. Hanaguri, C. Lupien, Y. Kohsaka, D.-H. Lee, M. Azuma, M. Takano, H. Takagi, J. C. Davis, Nature \textbf{430}, 1001 (2004); M. J. Lawler, K. Fujita, Jhinhwan Lee, A. R. Schmidt, Y. Kohsaka, Chung Koo Kim, H. Eisaki, S. Uchida, J. C. Davis, J. P. Sethna and Eun-Ah Kim, Nature \textbf{466}, 347 (2010). 

\bibitem{Varma} C.M. Varma, Phys. Rev. B \textbf{55} (1997) 14554; Phys. Rev.
Lett. \textbf{83} (1999) 3538.

\bibitem{Bourges} P. Bourges, Y. Sidis, arXiv 1101.1786.

\bibitem{Alloul-EPL2010} H. Alloul, F. Rullier-Albenque, B. Vignolle, D. Colson, and A. Forget, Europhys. Lett. \textbf{91}, 37005 (2010).

\bibitem{Skocpol} W.J. Skocpol and M. Tinkham, Rep. Prog. Phys. \textbf{38}, 1049 (1975).

\bibitem{Pourret-NPhys2006} A. Pourret, H. Aubin, J. Lesueur, C. A. Marrache-kikuchi, L. Berg\'e, L. Dumoulin, and K. Behnia, Nat Phys. \textbf{2}, 683 (2006).

\bibitem{Pourret-NewPhys2009} A Pourret, P Spathis, H Aubin and K Behnia, New J. Phys. \textbf{11}, 055071 (2009).

\bibitem{Larkin-Varlamov} For a review see A. Larkin and A.A. Varlamov, Theory of fluctuations in
superconductors (Oxford University Press), Oxford, 2005.

\bibitem{SCF-YBCO} R. Hopfengartner, B. Hensel, and G. Saemann-Ischenko, Phys. Rev. B 
\textbf{44}, 741 (1991).

\bibitem{SCF-Bi2212} P. Mandal et al. Physica C \textbf{169}, 43 (1990).

\bibitem{SCF-Tl} H.M. Duan et al., Phys. Rev. B \textbf{43}, 12925 (1991).

\bibitem{Cimberle} M.R. Cimberle et al., Phys. Rev. B \textbf{55}, R14745 (1997).

\bibitem{Lang} W. Lang, G. Heine, W. Kula, R. Sobolewski, Phys. Rev. B \textbf{51},9180 (1995).

\bibitem{Semba} K. Semba, T. Ishii, and A. Matsuda, Phys. Rev. Lett. \textbf{67}, 769 (1991) 

\bibitem{Holm} W. Holm, O. Rapp, C.N.L. Johnson, and U. Helmersson, Phys. Rev. B \textbf{52}, 3748 (1995).

\bibitem{Bouquet} F. Bouquet, L. Fruchter, I. Sfar, Z. Z. Li, and H. Raffy, Phys. Rev. B \textbf{74}, 064513 (2006).

\bibitem{Sekirnjak} C. Sekirnjak, W. Lang, S. Proyer, P. Schwab, Physica C \textbf{243}, 60 (1995).

\bibitem{Ando-PRL2002} Y. Ando and K. Segawa, Phys. Rev. Lett. \textbf{88}, 167005 (2002).

\bibitem{RA-HF} F.\ Rullier-Albenque, H. Alloul, C. Proust, D. Colson, and A. Forget, Phys. Rev. Lett. \textbf{99}, 027003 (2007).

\bibitem{Tallon} J.L. Tallon, C. Bernhard, H. Shaked, R.L. Hitterman, J.D. Jorgensen, Phys. Rev. B \textbf{51}, 12911 (1995).

\bibitem{RMP} H. Alloul, J. Bobroff, M. Gabay and P. Hirschfeld, Rev.Mod. Phys. \textbf{81}, 45
(2009).

\bibitem{Legris} A. Legris, F. Rullier-Albenque, E. Radeva and P. Lejay, J. Phys. I 
France \textbf{3}, 1605 (1993).

\bibitem{FRA-EPL} F. Rullier-Albenque, P.A. Vieillefond, H. Alloul, A.W. Tyler, P. Lejay, and J.F. Marucco,
Europhys. Lett. \textbf{50}, 81 (2000).

\bibitem{FRA-PRL2001} F. Rullier-Albenque, H. Alloul, R. Tourbot, Phys. Rev. Lett. \textbf{87}, 157001 (2001). 

\bibitem{FRA-PRL2003} F.Rullier-Albenque, H. Alloul, R. Tourbot, Phys. Rev. 
Lett. \textbf{91}, 047001 (2003).

\bibitem{Lacerda} A. Lacerda, J.P. Rodriguez, M.F. Hundley, Z. Fisk, P.C. Canfield , J.D. Thompson, S.W. Cheong, Phys. Rev. B \textbf{49}, 9097 (1994). 

\bibitem{Harris} J.M. Harris, Y.F. Yan, P. Matl, N.P. Ong, P.W. Anderson, T. Kimura, and K. Kitazawa, 
Phys. Rev. Lett. \textbf{75}, 1391 (1995).

\bibitem{Konstantinovic} Z. Konstantinovic, O. Laborde, P. Monceau, Z.Z. Li , H. Raffy, Physica B 
\textbf{259-261},569 (1999). 

\bibitem{Tyler-thesis} A.W. Tyler, Thesis, University of Cambridge (1997).

\bibitem{footnoteMR} For the underdoped sample, the longitudinal MR component has been shown to be significant at high temperature \cite{Harris}. We did not measure it but assume here that it remains quadratic in field up to 60T.
Any saturation of this MR would only marginally change the results.

\bibitem{Doiron-Leyraud} N. Doiron-Leyraud, C. Proust, D. LeBoeuf, J. Levallois,
J-B. Bonnemaison, R. Liang, D.A. Bonn, W.N. Hardy and L. Taillefer, Nature \textbf{447},565 (2007)

\bibitem{Leboeuf} D. LeBoeuf, N. Doiron-Leyraud, J. Levallois, R. Daou,
J.-B. Bonnemaison, N.E. Hussey, L. Balicas, B.J. Ramshaw, Ruixing Liang,
D.A. Bonn, W.N. Hardy, S. Adachi, C. Proust and L. Taillefer, Nature \textbf{450}, 533 (2007). 

\bibitem{footnotedatalowT} In the case of the OPT93.6 and UD57 samples, we have added data taken
in a range where the magnetic field is not actually large enough to totally 
suppress the superconductivity. At each $T$, we have used the corresponding value of $a_{trans}$ 
by extrapolating the curve of Fig.\ref{Fig.magneto-coeff} and calculated the curve of the normal 
state resistivity versus magnetic field with different values of $\rho_{n}$ for the highest applied 
magnetic field. This allows us to deduce values of $\Delta\sigma_{SF}(T,0)$ within the indicated error bars.

\bibitem{FRA-EPL-MIT} F. Rullier-Albenque, H. Alloul, F. Balakirev, C. Proust, Europhys. 
Lett. \textbf{81}, 37008 (2008).

\bibitem{Bergeal} N. Bergeal, J. Lesueur, M. Aprili, G. Faini, J.P. Contour, and B. Leridon, Nature Physics
\textbf{4}, 608 (2008).

\bibitem{Grbic-YBCO} M. S. Grbi´c, M. Pozek, D. Paar, V. Hinkov, M. Raichle, D. Haug, B. Keimer, 
N. Barisi´c, and A. Dulci´c, arXiv 1005.4789, unpublished.

\bibitem{Grbic-Hg} M.S. Gbric et al., Phys. Rev. B \textbf{80}, 094511 (2010).

\bibitem{Wang-science} Y. Wang, S. Ono, Y. Onose, G. Gu, Y. Ando, Y. Tokura, S. Uchida, N.P. Ong,
Science \textbf{299},86 (2003).

\bibitem{Li-M2S}L. Li, Y. Wang, J. G. Checkelsky, M. J. Naughton,
S. Komiya, S. Ono, Y. Ando and N. P. Ong, proc. M2S-HTSC-VIII, Dresden 2006, Physica C \textbf{460-462}, 48 (2007). 

\bibitem{Yip} S.K. Yip, Phys. Rev. B \textbf{41}, 2612 (1990).

\bibitem{Lawrence-Doniach} W.E. Lawrence and S. Doniach, proceedings 12th International Conference on Low
Temperature Physics, Kyoto 1970, edited by E. Kanda (Keigaku, Tokyo, (1971), p361.

\bibitem{Freitas} P.P. Freitas, C.C. Tsuei, and T.S. Plaskett, Phys. Rev. B \textbf{36}, 833 (1987).

\bibitem{Hopfengartner} R. Hopfeng\''artner, B. Hensel, and G. Saemann-Ischenko, Phys. Rev. B \textbf{44}, 741 (1991).

\bibitem{Gauzi} A. Gauzzi and D. Pavuna, Phys. Rev. B \textbf{51}, 15420 (1995).

\bibitem{Ussishkin} I. Ussishkin, S.L. Sondhi and D.A. Huse , Phys. Rev. Lett.
\textbf{89} 287001 (2002). 

\bibitem{Leridon-PRB2007} B. Leridon, J. Vanacken, T. Wambecq, and V. V. Moshchalkov, 
Phys. Rev. B textbf{76}, 012503 (2007).

\bibitem{footnote-Nernst} As explained in ref.\cite{FRA-Nernst}, the off-diagonal Peltier term $\alpha_{xy}$ is determined from the Nernst coefficient $\nu$, the Hall angle $\theta_{H}$, the conductivity $\sigma$ and the thermopower coefficient $S$ as $\alpha_{xy}=\sigma [\nu B+S\tan \theta_H]$. 

\bibitem{Daou} R. Daou, J. Chang, David LeBoeuf, Olivier Cyr-Choinière, Francis Laliberté, Nicolas Doiron-Leyraud, B. J. Ramshaw, Ruixing Liang, D. A. Bonn, W. N. Hardy, and L. Taillefer, Nature \textbf{463} (2010) 519. 

\bibitem{Kokanovic} I. Kokanovic, J.R. Cooper and M.Matusiak, Phys. Rev. Lett. \textbf{102}, 187002 (2009).

\bibitem{Lee-RMP} P.A. Lee, N. Nagaosa, X.G. Wen, Rev. Mod. Phys. \textbf{78}, 17 (2006).

\bibitem{Halperin-Nelson} B. I. Halperin and D. R. Nelson, J. Low. Temp. Phys. \textbf{36},
599 (1979).

\bibitem{Benfatto} L. Benfatto, C. Castellani, and T. Giamarchi, Phys. Rev. B \textbf{80}, 214506 (2009).

\bibitem{Tallon-2010} J.L. Tallon, J.G. Storey and J.W. Loram, ArXiv 0908.4428, unpublished.

\bibitem{Hikami} S. Hikami and A.I. Larkin, Mod. Phys. Lett. B \textbf{2}, 693 (1988).  

\bibitem{Matsuda} Y. Matsuda, T. Hirai, S. Komiyama, T. Terashina, Y. Bando, K. Iijima, K. Yamamoto, and K. Hirata, Phys. Rev. B \textbf{40}, 5176 (1989).

\bibitem{Kapitulnik} A. Kapitulnik, A. Palevski and G. Deutscher, J. Phys. C: Solid State Phys., \textbf{18} 1305 (1985).

\bibitem{Pourret-PRB2007}A. Pourret, H. Aubin, J. Lesueur, C. A. Marrache-Kikuchi, L. Berg\'e, L. Dumoulin, and K. Behnia, Phys. Rev. B, \textbf{76},214504 (2007).

\bibitem{Zhou} X. J. Zhou, T. Yoshida, A. Lanzara, P. V. Bogdanov, S. A. Kellar, K. M. Shen, W. L. Yang, F. Ronning, T. Sasagawa, T. Kakeshita,T. Noda, H. Eisaki, S. Uchida, C. T. Lin, F. Zhou, J. W. Xiong, W. X. Ti, Z. X. Zhao, A. Fujimori,
Z. Hussain, Z.-X. Shen, Nature \textbf{423}, 398 (2003).

\bibitem{Fournier} D. Fournier, G. Levy, Y. Pennec, J.L. McChesney, A. Bostwick, E. Rotenberg,
R. Liang, W.N. Hardy, D.A. Bonn, I.S. Elfimov, and A. Damascelli, Nature Physics \textbf{6}, 905 (2010).

\bibitem{Hufner} S H\"ufner, M A Hossain, A Damascelli and G A Sawatzky, Rep. Prog. Phys. \textbf{71}, 062501 (2008).

\bibitem{Yazdani} A. Yazdani, J. Phys.: Condens. Matter \textbf{21}, 164214 (2009).

\bibitem{Hudson} M. C. Boyer, W. D. Wise, K. Chatterjee, M. Yi, T. Kondo, T. Takeuchi, H. Ikuta and E. W. Hudson, 
Nature Physics \textbf{3}, 802 (2007).

\bibitem{Reggiani} L. Reggiani, R. Vaglio, and A.A. Varlamov, Phys. Rev B \textbf{44}, 9541 (1991).

\bibitem{Glatz-2010} A. Glatz, A. A. Varlamov, and V. M. Vinokur, arXiv 1012.1104, unpublished.

\bibitem{Balestrino} G. Balestrino, M. Marinelli, E. Milani, L. Reggiani, R. Vaglio,
and A. Varlamov, Phys. Rev. B \textbf{46}, 14919 (1992).

\bibitem{Carballeira} C. Carballeira, S.R. Curras, J. Vi\~na, J.A. Veiro, M.V. Ramallo, F. Vidal, Phys. Rev. B \textbf{63}, 144515 (2001).

\bibitem{Vidal} F. Vidal, C. Carballeira, S.R. Curras, J. Mosqueira, M.V. Ramallo, J.A. Veira and J. Vina, Europhys. Lett. \textbf{59}, 754 (2002).

\bibitem{Levchenko-2010} Alex Levchenko,M. R. Norman, and A. A. Varlamov, arXiv 1009.2213, unpublished.

\bibitem{McElroy} K. McElroy, D.-H. Lee, J. E. Hoffman, K. M. Lang, J. Lee, E.W. Hudson, H. Eisaki,
S. Uchida, and J. C. Davis, Phys. Rev. Lett.\textbf{94}, 197005 (2005).

\bibitem{LeTacon} M. Le Tacon, A. Sacuto, A. Georges, G. Kottliar, Y. Gallais, D. Colson, A. Forget, Nature Physics \textbf{2}, 537 (2006).

\bibitem{Kondo} T. Kondo, R. Khasanov, T. Takeuchi, J. Schmalian and A. Kaminski, Nature \textbf{457}, 296 (2009).

\bibitem{Kontani} H. Kontani, Phys. Rev. Lett. \textbf{96}, 067002 (2006).

\bibitem{Behnia-Nernst} K. Behnia, J. Phys.: Condens. Matter \textbf{21}, 113101 (2009).

\bibitem{Vojta} A. Hackl and M. Vojta, ArXiv 0909.4534 unpublished.

\bibitem{Pan} S. H. Pan, J. P. O'Neal, R. L. Badzey, C. Chamon, H. Ding,
J. R. Engelbrecht, Z. Wang, H. Eisaki, S. Uchida, A. K. Guptak,
K.-W. Ngk, E. W. Hudson, K. M. Lang and J. C. Davis, Nature \textbf{413}, 282 (2001).

\bibitem{Bobroff-PRL2002} J. Bobroff, H. Alloul, S. Ouazi, P. Mendels, A. Mahajan, N. Blanchard, G. Collin, V. Guillen, and J.-F. Marucco, Phys. Rev. Lett.\textbf{89}, 157002 (2002).

\bibitem{Fisher} \O. Fischer, M. Kugler, I. Maggio-Aprile, C. Berthod and C. Renner, Rev. Mod. Phys. \textbf{79}, 353 (2007).

\bibitem{Sacepe1}B. Sac\'ep\'e, C. Chapelier, T. I. Baturina, V. M. Vinokur, M. R. Baklanov, and M. Sanquer, Phys. Rev. Lett \textbf{101}, 157006 (2008); B. Sac\'ep\'e, T. Dubouchet, C. Chapelier, M.
Sanquer, M. Ovadia, D. Shahar, M. Feigel'man, and L. Ioffe, arXiv1012.3630 unpublished.

\bibitem{Sacepe2} B. Sac\'ep\'e, C. Chapelier, T. I. Baturina, V. M. Vinokur, M. R.
Baklanov, and Marc Sanquer, Nature Communications, \textbf{1}, 140 (2010).

\end{thebibliography}
\end{document}